%% file: higherd_arXiv2.tex
\documentclass[%
 reprint,
preprintnumbers,
nofootinbib,
 amsmath,amssymb,
 aps,
onecolumn
]{revtex4-2}

\usepackage{graphicx}
\usepackage{dcolumn}
\usepackage{bm}
\usepackage{hyperref}

\usepackage{ulem}
\usepackage{blkarray, multirow}

\def\mr{\mathrm}

\def\nm{\nonumber}
\def\pd{\partial}

\def\rmd{{\rm d}}

\def\umetric{{\mathfrak g}}

\def\pX{{\mathcal X}}
\def\pN{{\mathcal N}}

\def\att{{A^{\rm (t)}}} 
\def\ett{{\Gamma}} 
\def\ati{{A}} 
\def\cti{{C}} 
\def\as{{A^{\rm (s)}}} 
\def\al{{A^{\rm (l)}}} 
\def\cs{{C^{\rm (s)}}} 
\def\cl{{C^{\rm (l)}}} 
\def\aij{{A}} 
\def\cij{{C}} 
\def\atr{{A^{\rm (tr)}}}
\def\ass{{A^{\rm (ss)}}}
\def\asl{{A^{\rm (sl)}}}
\def\all{{A^{\rm (ll)}}}
\def\ctr{{C^{\rm (tr)}}}
\def\css{{C^{\rm (ss)}}}
\def\csl{{C^{\rm (sl)}}}
\def\cll{{C^{\rm (ll)}}}

\def\cu3{{\tilde C^{\rm (ll)}}}

\def\detU{{\mathcal U}}

\newcommand{\dif}[2]{\frac{\mr{d} #1}{\mr{d} #2}}
\newcommand{\hdif}[3]{\frac{\mr{d}^{#1} #2}{\mr{d} #3^{#1}}}
\newcommand{\pdif}[2]{\frac{\pd #1}{\pd #2}}
\newcommand{\hpdif}[3]{\frac{\pd^2 #1}{\pd #2 \pd #3}}

\begin{document}

\preprint{YITP-22-94}

\title{Ostrogradsky mode in scalar-tensor theories \\
with higher-order derivative couplings to matter}

\author{Atsushi Naruko}
 \affiliation{%
 Center for Gravitational Physics and Quantum Information, Yukawa Institute for Theoretical Physics, Kyoto University, Kyoto 606-8502, Japan
 }
 
\author{Ryo Saito}%
\affiliation{%
 Graduate School of Science and Engineering, Yamaguchi University, Yamaguchi 753-8512, Japan
}%
\affiliation{%
Kavli Institute for the Physics and Mathematics of the Universe, Todai Institute for Advanced Study, The University of Tokyo, Chiba 277-8583, Japan (Kavli IPMU, WPI)
}%

\author{Norihiro Tanahashi}
\affiliation{%
Department of Physics, Chuo University, Kasuga, Bunkyo-ku, Tokyo 112-8551, Japan
}%

\author{Daisuke Yamauchi}
\affiliation{%
Faculty of Engineering, Kanagawa University, Kanagawa 221-8686, Japan
}%

\date{\today}

\begin{abstract}
A metric transformation is a tool to find a new theory of gravity beyond general relativity. The gravity action is guaranteed to be free from a dangerous Ostrogradsky mode as long as the metric transformation is regular and invertible.
Various degenerate higher-order scalar-tensor theories (DHOST) without extra degrees of freedom have been found through the metric transformation with a scalar field and its derivatives.
In this work, we examine how a matter coupling changes the degeneracy for a theory generated from the Horndeski theory through the metric transformation with the second derivative of a scalar field, 
taking a minimally-coupled free scalar field as the matter field. 
When the transformation is invertible, 
this theory is equivalent to the Horndeski theory with a higher-order derivative coupling to the matter scalar field.
Working in this Horndeski frame and the unitary gauge, 
we find that the degeneracy conditions are solvable and the matter metric must have a certain structure to remove the Ostrogradsky mode.
\end{abstract}

\maketitle

\section{Introduction}
\label{s:int}

Ostrogradsky has shown that Hamiltonian is bounded neither below nor above if the equation of motion of a system without the degeneracy is higher than second order~\cite{Ostrogradsky:1850fid,Woodard:2015zca}. 
This result implies that there would be instability in the system. 
A scalar-tensor theory is a natural extension of the theory of gravity, general relativity, where a scalar degree of freedom is present 
in addition to the degree of freedom of gravitational waves.
The Ostrogradsky mode appears in this type of theory when the Lagrangian includes quadratic or even more terms of the second-order derivative of the scalar field unless some mechanisms to suppress the Ostrogradsky mode are introduced.
Recent studies revealed the existence of scalar-tensor theories, dubbed Degenerate Higher-order Scalar-Tensor (DHOST) theories, which are free from the Ostrogradsky mode thanks to the degeneracy while their Lagrangian include quadratic or even more terms of the second-order derivative of the scalar field 
\cite{Zumalacarregui:2013pma, Gleyzes:2014dya, Gleyzes:2014qga,  Langlois:2015cwa,Crisostomi:2016czh, BenAchour:2016cay, BenAchour:2016fzp}. 
Equations of motion of these theories 
are of higher than second order in general, and then 
these theories are clearly beyond the Horndeski theory 
\cite{Horndeski:1974wa, Deffayet:2011gz, Kobayashi:2011nu}, 
whose equations of motion are up to second order.
For a review of these recent developments of scalar-tensor theories, see Refs.~\cite{Langlois:2018dxi,Kobayashi:2019hrl}.
More recently, 
scalar-tensor theories with more than second-order derivatives in Lagrangian are discussed, 
focusing on metric transformations with higher derivatives of the scalar field \cite{Alinea:2020laa,Minamitsuji:2021dkf,
Babichev:2021bim,Babichev:2019twf,Takahashi:2021ttd} 
(see also Refs.~\cite{Motohashi:2017eya, Motohashi:2018pxg}).

A metric transformation is a tool to find a new {\it healthy} theory of gravity beyond general relativity free from the Ostrogradsky mode. 
The first example of theories beyond Horndeski was found in Ref.~\cite{Zumalacarregui:2013pma} via a metric transformation of the Einstein-Hilbert action. 
It was studied in Refs.~\cite{Bettoni:2013diz, Gleyzes:2014dya, Gleyzes:2014qga, Crisostomi:2016czh, BenAchour:2016cay, BenAchour:2016fzp} how different DHOST theories are related via the conformal and disformal transformations \cite{Bekenstein:1992pj}.
For example, the so-called DHOST class I theory is shown to be equivalent to the Horndeski theory with a metric transformation. 
Further extension of the DHOST theories along this direction was made in Refs.~\cite{Alinea:2020laa,Minamitsuji:2021dkf,
Babichev:2021bim,Babichev:2019twf,Takahashi:2021ttd} with a higher-derivative generalization of the disformal transformation.
As long as the metric transformation is regular and invertible, 
the metric transformation
is nothing but a rewriting of physical metric in a different manner and hence the physics is essentially the same \cite{Deruelle:2010ht, Deruelle:2014zza, Arroja:2015wpa, Domenech:2015tca, Takahashi:2017zgr}
(see Ref.~\cite{Jirousek:2022jhh} for a recent discussion on exceptional cases).
Therefore, the new theory is guaranteed to be healthy if the original theory is healthy.
However, in the presence of matter, the situation 
completely changes. 
When the new metric is assumed to be minimally coupled 
to matter fields, the new theory is no longer equivalent to the original theory with minimally-coupled matter fields. 
Then, one may wonder whether the tamed Ostrogradsky mode due to the degeneracy may revive once the matter coupling is taken into account~\cite{Deffayet:2020ypa}.

Provided the weak equivalence principle holds, there should be a metric universally coupled to matter fields. 
Then the natural question would be ``to which metric the matter fields are minimally coupled?" 
One can consider the gravity theory described by a metric $g_{\mu\nu}$ and matter fields $\sigma_a$ minimally coupled to this metric, which we dub $G$(minimal):
    \begin{align}
        S^{G\text{(minimal)}}[g_{\mu\nu},\sigma_a] = \int \rmd^4 x \sqrt{- g} \Bigl( {\cal L}^G [g_{\mu\nu}] + {\cal L}^m [g_{\mu\nu} \,, \sigma_a] \Bigr) \,.
    \end{align}
Another gravity theory $\overline{{\cal L}}{}^G [g_{\mu\nu}]$ can be generated through a metric transformation ${\cal T}: g_{\mu\nu} \to \bar{g}_{\mu\nu}$ as 
    \begin{align}
        \sqrt{-g}\, \overline{{\cal L}}{}^G [g_{\mu\nu}]=\sqrt{-\bar{g}}{\cal L}^G [\bar{g}_{\mu\nu}] \,; \quad \bar{g}_{\mu\nu} = {\cal T}[g_{\mu\nu}] \,.
    \end{align}
Let us first assume that the matter fields are coupled to the metric $\bar{g}_{\mu\nu}$. 
As a theory for the metric $g_{\mu\nu}$, 
this new (modified) gravity theory has a non-minimal coupling to matter. 
We call this theory $\overline{G}$(non-minimal):
    \begin{align}
        S^{\overline{G}(\text{non-minimal})}[g_{\mu\nu},\sigma_a] = \int \rmd^4 x \Bigl( \sqrt{- g}\,\overline{\cal L}{}^G [g_{\mu\nu}] + \sqrt{- \bar{g}}{\cal L}^m [\bar{g}_{\mu\nu} \,, \sigma_a] \Bigr) \,; \quad \bar{g}_{\mu\nu} = {\cal T}[g_{\mu\nu}] \,.
    \end{align}
However, 
when the metric $\bar{g}_{\mu\nu}$ is considered to be an independent variable, 
the $\overline{G}$(non-minimal) theory simply reduces to $G$(minimal): 
    \begin{align}
         S^{\overline{G}\text{(non-minimal)}}[g_{\mu\nu},\sigma_a] = \int \rmd^4 x \Bigl( \sqrt{- \bar{g}}{\cal L}^G [\bar{g}_{\mu\nu}] + \sqrt{-\bar{g}}{\cal L}^m [\bar{g}_{\mu\nu} \,, \sigma_a] \Bigr) 
         = S^{G\#\text{minimal}}[\bar{g}_{\mu\nu},\sigma_a] \,.
    \end{align}
One can also consider a theory where
the gravity theory is given by $\overline{{\cal L}}{}^G [g_{\mu\nu}]$ while the matter fields are coupled to the metric $g_{\mu\nu}$, which we call $\overline{G}$(minimal):
    \begin{align}   \label{caseC1}
        S^{\overline{G}\text{(minimal)}}[g_{\mu\nu},\sigma_a]
        = \int \rmd^4 x \sqrt{- g} \Bigl( \overline{\cal L}{}^G [g_{\mu\nu}] + {\cal L}^m [g_{\mu\nu} \,, \sigma_a] \Bigr) \,.
    \end{align}
As a theory for the metric $g_{\mu\nu}$, 
the new gravity theory has a minimal coupling to matter.
When the metric $\bar{g}_{\mu\nu}$ is considered to be an independent variable, 
this theory reduces to the original theory but the matter is now non-minimally coupled:
    \begin{align}   \label{caseC2}
        S^{\overline{G}\text{(minimal)}}[g_{\mu\nu},\sigma_a] = \int \rmd^4 x  \Bigl( \sqrt{- \bar{g}}{\cal L}^G [\bar{g}_{\mu\nu}] + \sqrt{- g}{\cal L}^m [g_{\mu\nu} \,, \sigma_a] \Bigr) 
        = S^{G\text{(non-minimal)}}[\bar{g}_{\mu\nu}, \sigma_a] \,.
    \end{align}
In the last expression, $g_{\mu\nu}$ in the middle equation is considered as a function of $\bar{g}_{\mu\nu}$ through the inverse transformation: $g_{\mu\nu} = {\cal T}^{-1}[\bar{g}_{\mu\nu}]$. 
The $\overline{G}$(minimal) theory is different from $G$(minimal) as well as $\overline{G}$(non-minimal). 
The matter metric $\bar{g}_{\mu\nu}$ contains higher derivatives when we consider the metric transformation with higher derivatives. In this case, the last expression in Eq.~(\ref{caseC2}) implies that, in the presence of matter, 
the theory is non-degenerate in general and then so is the equivalent theory (\ref{caseC1}), 
while the gravity sector $\overline{\cal L}{}^G [g_{\mu\nu}]$ satisfies the degeneracy condition and the matter sector does not contain higher derivatives. 
Even if the gravity sector itself is well-behaved, once we take into account a matter sector, a careful analysis is needed since the Ostrogradsky mode (higher-derivative mode) can revive through the matter coupling \cite{Deffayet:2020ypa}.

In this paper, we investigate how the matter coupling changes the degeneracy of a theory generated from the Horndeski theory through the metric transformation ${\cal T}: g_{\mu\nu} \to \bar{g}_{\mu\nu}$ with
    \begin{align}   \label{bargdef}
		\bar{g}_{\mu\nu} = F_0(\phi, X,Y,Z) g_{\mu\nu} + F_1(\phi, X,Y,Z) \phi_\mu \phi_\nu + 2F_2(\phi, X,Y,Z) \phi_{(\mu}X_{\nu)} + F_3(\phi, X,Y,Z) X_\mu X_\nu~,
	\end{align}
which contains
the second derivatives of a scalar field $\phi$. Here, we defined
	\begin{align}
		X \equiv \phi_\mu \phi^\mu \,, \quad
		Y \equiv \phi_\mu X^\mu \,, \quad
		Z \equiv X_\mu X^\mu \,,
	\end{align}
and $\phi_\mu \equiv \nabla_\mu \phi$ and $X_\mu \equiv \nabla_\mu X$.
This is the metric transformation studied in Ref.~\cite{Takahashi:2021ttd} where the invertibility conditions of the transformation are derived for the functions $F_a~(a=0,1,2,3)$. 
For $F_0=F_0(\phi, X)$, $F_1=F_1(\phi,X)$, and $F_2=F_3=0$, the transformation is known as a disformal transformation \cite{Bekenstein:1992pj}. 
Taking a free scalar field as the matter sector, 
we will show that the degeneracy conditions are solvable in the unitary gauge $\phi=t$ and restrict the form of the transformation (\ref{bargdef}) as follows:
    \begin{align}   \label{eq:resultant transformation extended intro}
	    \bar g_{\mu\nu}=F_\phi^{\rm U}(\phi ,X, W)\phi_\mu\phi_\nu 
     + F_{\parallel}^{\rm U}(\phi ,X,W) \left( X_\mu - V^{\rm U} \phi_\mu \right) \left( X_\nu - V^{\rm U} \phi_\nu \right)
     + XW F^{\rm U}_{\perp}(\phi,X,W)\gamma^{(\perp)}_{\mu\nu} 
	\,,
    \end{align}
where
    \begin{align}
        \pX_\mu \equiv X_\mu - \frac{Y}{X}\phi_\mu \quad  ( \pX_\mu \pX^\mu = W ) \,, \quad 
        \gamma^{(\perp)}_{\mu\nu} = g_{\mu\nu} - \frac{\phi_\mu\phi_\nu}{X} - \frac{ \pX_\mu \pX_\nu}{W} \,,
    \end{align}
and $V^{\rm U}_\parallel=V^{\rm U}_\parallel(\phi, X, Y, W)$ is linear in $Y$. 
The functions $F_a^{\rm U}(\phi,X,W)~(a=\phi, \parallel, \perp)$ and $V^{\rm U}(\phi, X, Y, W)$ can be chosen independently. 
They are related to the functions $F_a(\phi,X,Y,Z)$ in Eq.~(\ref{bargdef}) as
    \begin{align}\label{eq:resultant restrictions extended intro}
        \begin{aligned}
            &F_0(\phi,X,Y,Z)= XW F_\perp^{\rm U}(\phi, X,W)
            \,,\ \ 
            F_1(\phi,X,Y,Z)=F_\phi^{\rm U}(\phi ,X,W) + [V^{\rm U}]^2 F_\parallel^{\rm U}(\phi ,X,W) - Z F_\perp^{\rm U}(\phi,X,W)
            \,,\\
            &F_2(\phi,X,Y,Z)=-V^{\rm U} F_\parallel^{\rm U}(\phi, X, W) + Y F_\perp^{\rm U}(\phi,X,W)
            \,,\ \ 
            F_3(\phi,X,Y,Z)=F_\parallel^{\rm U}(\phi, X,W) - X F_\perp^{\rm U}(\phi,X,W)
            \,.
        \end{aligned}
    \end{align}
The dependence on the higher-derivative terms $Y$ and $Z$ 
in the functions $F_a~(a=0,1,2,3)$ is tightly restricted as the above relations to achieve the cancellation between the higher-derivative terms in the metric (\ref{eq:resultant transformation extended intro}). 
Therefore, the matter coupling changes the degeneracy of the theory. 

The paper is organized as follows: 
In Section \ref{s:models}, we define the models analyzed in this paper and their expression in the unitary gauge. 
In Section \ref{s:dc}, we show how the degeneracy conditions restrict the form of the metric (\ref{bargdef}). 
We then derive the restrictions when the invertibility conditions are further imposed in Section \ref{s:invertible} and give the covariant form of the resultant restricted metric in Section \ref{s:covariant}. 
We finally summarize the results in Section \ref{s:summary}. 
Some useful expressions for the computation are presented in two Appendices \ref{a:FUrelations} and \ref{a:det}. 
In Appendix \ref{a:frames}, we discuss how the structure of degeneracy changes in different frames for a simple example of a mechanical system. 

\section{Models}
\label{s:models}

\subsection{Action}
\label{ss:action}

We consider the following action for the Horndeski theory with a non-minimal coupling to a matter scalar field $\sigma$,
	\begin{align}   \label{Lag}
		 S^{H\text{(non-minimal)}}[g_{\mu\nu},\phi,\sigma] = \int \rmd^4 x \Bigl[ \sqrt{-g} {\cal L}^{\rm H}[g_{\mu\nu}, \phi]+ \sqrt{-\bar{g}}{\cal L}^{\rm m}[\bar{g}_{\mu\nu}, \sigma] \Bigr] \,,
	\end{align}
where the matter metric $\bar{g}_{\mu\nu}$ is assumed to be a function of $g_{\mu\nu}$ and $\phi$ in the form (\ref{bargdef}),
    \begin{align}   \label{bargdef2}
		\bar{g}_{\mu\nu} = F_0(\phi, X,Y,Z) g_{\mu\nu} + F_1(\phi, X,Y,Z) \phi_\mu \phi_\nu + 2F_2(\phi, X,Y,Z) \phi_{(\mu}X_{\nu)} + F_3(\phi, X,Y,Z) X_\mu X_\nu 
		\,.
	\end{align}
Here, ${\cal L}^{\rm H}$ and ${\cal L}^{\rm m}$ represent the Lagrangian densities of the Horndeski theory and matter, respectively. 
When the transformation ${\cal T}: g_{\mu\nu} \to \bar{g}_{\mu\nu}$ is invertible, 
this theory is equivalent to any new theory of gravity generated from the Horndeski action through the metric transformation in the form (\ref{bargdef2}) with a minimal coupling to the matter scalar field $\sigma$:
provided the inverse transformation ${\cal T}^{-1}: g_{\mu\nu} \to \hat{g}_{\mu\nu}$, we find
	\begin{align}\label{barH}
		 S^{H\text{(non-minimal)}}[\hat{g}_{\mu\nu},\phi,\sigma] = \int \rmd^4 x \Bigl[ \sqrt{-\hat{g}} {\cal L}^{\rm H}[\hat{g}_{\mu\nu}, \phi]+ \sqrt{-g}{\cal L}^{\rm m}[g_{\mu\nu}, \sigma] \Bigr] = S^{\overline{H}\text{(minimal)}}[g_{\mu\nu},\sigma]\,.
	\end{align}
In the last expression, $\hat{g}_{\mu\nu}$ in the middle equation is considered as a function of $g_{\mu\nu}$: $\hat{g}_{\mu\nu} = {\cal T}^{-1}[g_{\mu\nu}]$. 
Thus, the action (\ref{Lag}) incorporates the DHOST class I theories and their generalizations proposed in Ref.~\cite{Takahashi:2021ttd}. 
We investigate whether there exists a matter metric $\bar{g}_{\mu\nu}$ for which the new theory of gravity (\ref{Lag}) is healthy, other than the trivial case $\bar{g}_{\mu\nu}=g_{\mu\nu}$. 

For simplicity, we assume that the matter field is given by a free scalar field
	\begin{align}   \label{Lmatter}
		{\cal L}^{\rm m}(\bar{g}_{\mu\nu}, \sigma) = \frac{\bar{g}^{\mu\nu}}{2}\nabla_\mu \sigma \nabla_\nu \sigma \,.
	\end{align}
Introducing
	\begin{align}	\label{eq:umetric def}
		\umetric^{\mu\nu} \equiv \sqrt{-\bar{g}}\bar{g}^{\mu\nu} \,,
	\end{align}
the matter sector depends on the metric $\bar{g}_{\mu\nu}$ only through $\umetric^{\mu\nu}$:
    \begin{align}
        S^{\rm m} = \int\rmd^4x ~ \frac{\umetric^{\mu\nu}}{2}\nabla_\mu \sigma \nabla_\nu \sigma \,.
    \end{align}
%

\subsection{The matter metric in the unitary gauge}
\label{ss:ug}

To analyze the degeneracy conditions of the theory (\ref{Lag}), we work in the so-called unitary gauge, where the clock is synchronized with the scalar field $\phi$,
	\begin{align}	\label{unitarygauge}
		\phi = t\,,
	\end{align}
and the ADM decomposition of the metric $g_{\mu\nu}$,
    \begin{align}
        g_{\mu\nu}\rmd x^\mu \rmd x^\nu = - N^2 \rmd t^2 + \gamma_{ij}(\rmd x^i + N^i \rmd t)(\rmd x^j + N^i \rmd t) \,.
    \end{align}
This means that we will consider the U-degenerate conditions in Ref.~\cite{DeFelice:2018ewo}. 
The covariant degeneracy conditions are not necessarily to be imposed to eliminate the Ostrogradsky mode. 
As we shall see in the following, the degeneracy conditions in the unitary gauge are much simpler than the covariant one in the theory (\ref{Lag}).

In the unitary gauge, $\phi_\mu$ and $X$ are given by
	\begin{align}
		\phi_\mu &= \nabla_\mu t = -\frac{n_\mu}{N}\,,
		    \label{eq:nabla_mu phi unitary gauge}
		\\
		X &= -\frac{1}{N^2}\,,
		    \label{eq:X unitary gauge}
	\end{align}
where $n^\mu$ is the unit normal of a time-constant hypersurface.
Then, we find that $X_\mu$, $Y$, and $Z$ are given by
	\begin{align}
	    X_\mu &= \frac{2\nabla_\mu N}{N^3} \,, \\
		Y &= \frac{2 \nabla^\mu t \nabla_\mu N}{N^3} = 
		-\frac{2}{N^5}( \dot{N} - N^i \nabla_i N ) \,, \\
		Z &= \frac{4\nabla^\mu N \nabla_\mu N}{N^6} = 
		\frac{4}{N^6}\left[ -\frac{( \dot{N} - N^i \nabla_i N )^2}{N^2} + \gamma^{ij}\nabla_i N \nabla_j N \right]\,.
	\end{align}
Using these relations, $\umetric^{\mu\nu}$ in Eq.~(\ref{eq:umetric def}) can be expressed as the form
	\begin{align}   \label{eq:umetric}
		\umetric^{\mu\nu} = U_0 g^{\mu\nu} + U_1n^\mu n^\nu + 2U_2 n^{(\mu}\nabla^{\nu)}N + U_3 \nabla^\mu N \nabla^\nu N \,,
	\end{align}
where $U_a$ ($a=0,1,2,3$) are functions of the following spatial scalars:
	\begin{align}	\label{eq:variable def}
		N\,, \quad \rho \equiv \nabla^0 N = -\frac{\dot{N}-N^i \nabla_i N}{N^2} \,, \quad X_s \equiv \gamma^{ij}\nabla_i N \nabla_j N \,.
	\end{align}
These three scalar quantities are related to $X,Y,Z$ as
    \begin{align}
        N = \frac{1}{\sqrt{-X}} \,, \quad \rho = \frac{Y}{2(-X)^{\frac{3}{2}}} \,, 
        \quad 
        X_s = \frac{Y^2-XZ}{4X^4}\,.
    \end{align}
The explicit relations between the functions $U_a$ and $F_a$ are shown in Appendix \ref{a:FUrelations}. 
For later convenience, we also give explicit forms of the metric components:
    \begin{align}
		\umetric^{00} &= -V_0 
	    	\label{hatg00}
		\\
		\umetric^{0i} &= V_0 N^i + V_1 D^i N
		    \label{hatg0i}\\
		\umetric^{ij} &= U_0 \gamma^{ij} - V_0 N^iN^j - 2V_1N^{(i} D^{j)}N + U_3 D^i N D^j N\,,
		    \label{hatgij}
	\end{align}
where
we introduced new functions
    \begin{align}
		V_0 &\equiv \frac{U_0 - U_1 -2U_2N\rho - U_3 (N\rho)^2}{N^2} \,, 
		    \label{V0def} \\
		V_1 &\equiv \frac{U_2+U_3 N\rho}{N} \,, 
		    \label{V1def}
	\end{align}
and new variables $D^i N \equiv \gamma^{ij}\nabla_j N$. 
In Eqs.~(\ref{hatg0i}) and (\ref{hatgij}), 
we used the fact that $\nabla^i N$ is decomposed as $\nabla^i N = g^{i\mu}\nabla_\mu N = -\rho N^i + D^i N$.
As we will see, it is convenient to introduce the vector
    \begin{align}
	    \pN^\mu \equiv \nabla^\mu N - N\rho\, n^\mu = (g^{\mu\nu}+n^\mu n^\nu)\nabla_\nu N \,,
    \end{align}
which satisfies $n_\mu \pN^\mu=0$ and $\pN_\mu \pN^\mu = X_s$. 
Then, the expression (\ref{eq:umetric}) for $\umetric^{\mu\nu}$ can be written as
	\begin{align}   \label{eq:umetric pN}
		\umetric^{\mu\nu} = U_0 (g^{\mu\nu} + n^\mu n^\nu) - N^2 V_0 n^\mu n^\nu + 2N V_1 n^{(\mu}\pN^{\nu)} + U_3 \pN^\mu \pN^\nu \,,
	\end{align}
in terms of the functions $V_0$ and $V_1$ defined above as well as $U_0$ and $U_3$. 

\section{Degeneracy conditions}
\label{s:dc}

\subsection{Kinetic structure of the action}
\label{ss:ks}

In the Lagrangian density (\ref{Lag}),
the Horndeski term $\sqrt{-g} {\cal L}_{\rm Horndeski}$ does not contain $\dot{N}$ nor $\dot{\sigma}$ \cite{Langlois:2018dxi, Kobayashi:2019hrl} and the matter term $\sqrt{-\bar{g}}{\cal L}_{\rm matter}$ contains only $\dot{N}$ and $\dot{\sigma}$ as the time-derivative terms. 
Hence, the kinetic matrix of the theory (\ref{Lag}) is given by a block diagonal matrix composed of (i) the kinetic matrix of the Horndeski term and (ii) the kinetic matrix for $\dot N$ and $\dot \sigma$ obtained from the matter term:
    \begin{align}
         \begin{blockarray}{ccc}
             & & 
             \begin{matrix}
                \scalebox{0.8}{$\dot{N}$} & \scalebox{0.8}{$\dot{\sigma}$}         
             \end{matrix}
             \\
            \begin{block}{c(c|c)}
                & \ast & \\
                \cline{2-3}
                \begin{matrix} 
                    \scalebox{0.8}{$\dot{N}$} \\ 
                    \scalebox{0.8}{$\dot{\sigma}$} 
                \end{matrix}
                ~~ 
                &  & K_{ab}
                \\
            \end{block}
        \end{blockarray}
        \qquad \quad (a,b=N, \sigma) \,,
    \end{align}
where 
    \begin{align}   \label{def:K}
		K_{ab} \equiv \hpdif{(\sqrt{-\bar{g}}{\cal L}^{\rm m})}{\dot{q}^a}{\dot{q}^b} \qquad (q^N = N\,, q^\sigma = \sigma) \,,
	\end{align}
for the matter Lagrangian (\ref{Lmatter}). 
This simple form of the kinetic matrix is an advantage of working in the unitary gauge (\ref{unitarygauge}). 
In the covariant gauge, 
both the Horndeski term and the matter term depend on $\ddot{\phi}$ and thus the kinetic matrix has a more complicated structure.

The structure of the kinetic matrix changes due to the matter term as
    \begin{align}
        \begin{blockarray}{ccc}
             & & \scalebox{0.8}{$\dot{N}$} \\
            \begin{block}{c(c|c)}
                & \ast &  
                \\ 
                \cline{2-3}
                \scalebox{0.8}{$\dot{N}$}~~ & & 0
                \\
            \end{block}
        \end{blockarray}
        \qquad \quad
        \rightarrow
        \qquad
         \begin{blockarray}{ccc}
             & & 
             \begin{matrix}
                \scalebox{0.8}{$\dot{N}$} & \scalebox{0.8}{$\dot{\sigma}$}         
             \end{matrix}
             \\
            \begin{block}{c(c|c)}
                & \ast & \\
                \cline{2-3}
                \begin{matrix} 
                    \scalebox{0.8}{$\dot{N}$} \\ 
                    \scalebox{0.8}{$\dot{\sigma}$} 
                \end{matrix}
                ~~ 
                &  & K_{ab}
                \\
            \end{block}
        \end{blockarray}
        \qquad \,.
    \end{align}
This shows that the primary constraint $\pi_N \equiv {\pd (\sqrt{-g}{\cal L})}/{\pd \dot{N}} = 0$ in the Horndeski theory is lost due to the matter term.  
To keep the number of primary constraints, the reduced kinetic matrix (\ref{def:K}) should be degenerate: 
when there is a primary constraint between the conjugate momenta of $N$ and $\sigma$, 
	\begin{align}
		\Phi(\pi_N, \pi_\sigma) = 0 \qquad (\pi_N= {\pd (\sqrt{-g}{\cal L})}/{\pd \dot{N}} \,,\, 
		\pi_\sigma={\pd (\sqrt{-g}{\cal L})}/{\pd \dot{\sigma}})\,,
	\end{align}
this implies
	\begin{align}
		\pdif{\Phi}{\pi_a} \delta \pi_a = \pdif{\Phi}{\pi_a} K_{ab} \delta \dot{q}^b = 0 \,,
	\end{align}
for arbitrary $\delta \dot{q}^b$. 
Hence, 
	\begin{align}   \label{eq:eigen}
		\pdif{\Phi}{\pi_a} K_{ab} = 0 \,,
	\end{align}
must be satisfied. Therefore, the matrix (\ref{def:K}) has a zero eigenmode. 
Moreover, we require $\pd \Phi/\pd \pi_N \neq 0$ because the constraint function $\Phi$ must depend on $\pi_N$ to remove the would-be Ostrogradsky mode. 
We also require that the matter scalar field $\sigma$ is dynamical and thus ${\rm rank}(K) = 1$.

Using the fact that $\umetric^{\mu\nu}$ does not depend on $\dot \sigma$, 
Eq.~(\ref{def:K}) is evaluated as
\footnote{We thank T. Ikeda, K. Takahashi, and T. Kobayashi for pointing out a typo in the following equation in the first draft.}
	\begin{align}   \label{Kmatrix}
		K = 
		\begin{pmatrix}
			\frac{1}{2}(\pd_{\dot{N}}^2 \umetric^{\mu\nu}) \nabla_\mu \sigma \nabla_\nu \sigma & (\pd_{\dot{N}}\umetric^{0\mu}) \nabla_\mu \sigma \\
			(\pd_{\dot{N}}\umetric^{0\mu}) \nabla_\mu \sigma & \umetric^{00}
		\end{pmatrix}
		\,.
	\end{align}
Its determinant is given by
	\begin{align}   \label{eq:dk}
		D_K \equiv \frac{1}{2}\left[ \umetric^{00}(\pd_{\dot{N}}^2 \umetric^{\mu\nu}) - 2(\pd_{\dot{N}}\umetric^{0\mu})(\pd_{\dot{N}}\umetric^{0\nu}) \right] \nabla_\mu \sigma \nabla_\nu \sigma\,,
	\end{align}
and the degeneracy condition is
    \begin{align}   \label{eq:dc}
		D_K = 0 \,.
	\end{align}
The conditions $\pd \Phi/\pd \pi_N \neq 0$ and ${\rm rank}(K)=1$ require that $\pd \Phi/\pd \pi_a \propto \delta_{a\sigma}$ should not be the zero eigenmode in Eq.~(\ref{eq:eigen}).
From Eq.~(\ref{Kmatrix}), this requirement is satisfied if and only if
    \begin{align}   \label{eq:g00req}
        \umetric^{00} \neq 0 \,.
    \end{align}

It is noteworthy that the analysis would become complicated if we did not use the Horndeski frame (\ref{Lag}) but the new gravity frame (\ref{barH}) ({\it i.e.}, the minimal coupling frame for the invertible case). 
In the Horndeski frame (\ref{Lag}), it is apparent that the primary constraint $\pi_N \equiv {\pd (\sqrt{-g}{\cal L})}/{\pd \dot{N}} = 0$ for the Horndeski theory is lost due to the matter coupling. 
Therefore, the Ostrogradsky mode appears unless the functions $F_a$ are appropriately chosen. 
On the other hand, in the new gravity frame (\ref{barH}), the number of the primary constraints is unchanged because the matter term does not contain the derivative of the gravity fields. 
These facts suggest that the matter term changes the higher-order (secondary, tertiary,...) constraints in the new gravity frame.
To find the same restrictions, we would need to perform the full constraint analysis not only the kinetic matrix (see Appendix \ref{a:frames}). 

\subsection{Solutions of the degeneracy conditions}
\label{ss:dc}

As usual, 
we assume that the degeneracy condition (\ref{eq:dc}) is satisfied for all configurations. 
This is equivalent to requiring that the unwanted Ostrogradsky mode should not appear for any configuration in the theory.%
\footnote{To be exact, our analysis is only applied to a time-like configuration of $\phi$ because we are working in the unitary gauge (\ref{unitarygauge}).} 
Under this assumption, Eq.~(\ref{eq:dc}) is satisfied for arbitrary $\nabla_\mu \sigma$ and thus
	\begin{align}   \label{det0eq}
		\umetric^{00}(\pd_{\dot{N}}^2 \umetric^{\mu\nu}) - 2(\pd_{\dot{N}}\umetric^{0\mu})(\pd_{\dot{N}}\umetric^{0\nu}) = 0 
	\end{align}
must be identically satisfied.
Because this equation is satisfied for any value of $\dot{N}$, we can consider it as a differential equation for $\dot N$.
As shown in Eqs.~(\ref{hatg00})-(\ref{hatgij}), 
the components of $\umetric^{\mu\nu}$ depend on $\dot N$ only through $\rho$ defined in Eq.~(\ref{eq:variable def}).
Therefore, Eq.~(\ref{det0eq}) is equivalent to
	\begin{align}   \label{det0eq_2}
		\umetric^{00}(\pd_{\rho}^2 \umetric^{\mu\nu}) - 2(\pd_{\rho}\umetric^{0\mu})(\pd_{\rho}\umetric^{0\nu}) = 0\,,
	\end{align}
whose independent variables are $\rho$ as well as the scalars $N$ and $X_s$ in Eq.~(\ref{eq:variable def}). 
In the following, we will see how this equation restricts the forms of the functions $U_a~(a=0,1,2,3)$ in Eq.~(\ref{eq:umetric}).

\subsubsection{Restrictions on $\umetric^{\mu\nu}$}

We solve each component of Eq.~(\ref{det0eq_2}) as follows:

\begin{itemize}

\item $(\mu,\nu)=(0,0)$:

The $00$ component of Eq.~(\ref{det0eq_2}) is given by
	\begin{align}   \label{det0eq_00}
		\umetric^{00}(\pd_{\rho}^2 \umetric^{00}) - 2(\pd_{\rho}\umetric^{00})(\pd_{\rho}\umetric^{00}) = 0\,,
	\end{align}
and it can be solved for $\umetric^{00}$ as a differential equation for $\rho$. 
Recasting Eq.~(\ref{det0eq_00}) as
	\begin{align}
        \pd_{\rho}\left\{ [\umetric^{00}]^{-2}(\pd_{\rho} \umetric^{00}) \right\} = 0 \,,
	\end{align}
and we can easily integrate it as
	\begin{align}   \label{hatg00sol}
        \umetric^{00} = -\frac{\att}{1 + \ett \rho}
        \,,
	\end{align}
where $\ett\,,\att$ are integration constants depending on $N$ and $X_s$. 
All the integration constants below are also functions of $N$ and $X_s$ unless explicitly stated. 
A special solution $1/\rho$ is expressed as the limit $|\att| \to \infty $ with $\ett/\att \to \text{const}$.
$\att \neq 0$ follows from $\hat g^{00} \neq 0$.

\item $(\mu,\nu)=(0,i)$:

The $0i$ component of Eq.~(\ref{det0eq_2}) is given by
	\begin{align}
		\umetric^{00}(\pd_{\rho}^2 \umetric^{0i}) - 2(\pd_{\rho}\umetric^{00})(\pd_{\rho}\umetric^{0i}) = 0 \,.
	\end{align}
By substituting Eq.~(\ref{hatg00sol}), 
this equation can be integrated once as
    \begin{align}
        \pd_{\rho}\umetric^{0i} = \frac{\ati^i}{(1+\Gamma \rho)^2} \,, 
    \end{align}
where $\ati^i$ is an integration constant. 
Further integration gives
	\begin{align}\label{g0i sol}
		\umetric^{0i}= 
             \begin{cases}
		        \ati^i \rho + \cti^i & (\ett = 0) \\
		        -\frac{\ati^i}{\ett (1 + \ett \rho)}+ \cti^i & (\ett \neq 0)
		    \end{cases} 
		\,,
	\end{align}
where $\cti^i$ is an integration constant. 
From the spatial covariance, 
the integration constants $\ati^i \,, \cti^i$ should be decomposed as
	\begin{align}   \label{C23}
		\ati^i = \as N^i + \al D^i N \,,
		\qquad
		\cti^i = \cs N^i + \cl D^i N \,.
	\end{align}
Here, the coefficients $\as,\al,\cs,\cl$ are the functions of $N$ and $X_s$. 
Their superscripts indicate that the coefficients with ${\rm (s)}$ and ${\rm (l)}$ correspond to $N^i$ (shift) and $D^i N$ (lapse), respectively.

\item $(\mu,\nu)=(i,j)$:

The $ij$ component of Eq.~(\ref{det0eq_2}) is given by
	\begin{align}
		\umetric^{00}(\pd_{\rho}^2 \umetric^{ij}) - 2(\pd_{\rho}\umetric^{0i})(\pd_{\rho}\umetric^{0j}) = 0 \,.
	\end{align}
For any value of $\ett$, this equation becomes
	\begin{align}
            \pd_{\rho}^2 \umetric^{ij} 
            = -\frac{2A^i A^j}{\att ( 1 + \ett \rho)^3}
        \,,
	\end{align}
and its solution is given by
	\begin{align}\label{sol gij}
		\umetric^{ij} = 
              \begin{cases}
		        -\frac{\ati^i \ati^j}{\att}\rho^2 +  \aij^{ij} \rho + \cij^{ij}  & (\ett = 0) \\
		        -\frac{\ati^i \ati^j}{\att \ett^2 (1 + \ett \rho)}  + \aij^{ij} \rho + \cij^{ij} &  (\ett \neq 0) \\
		    \end{cases} 
		\,.
	\end{align}
Here, $\aij^{ij}\,, \cij^{ij}$ are integration constants and can be decomposed as
	\begin{align}   \label{C45}
	    \begin{aligned}
	        \aij^{ij} &= \atr \gamma^{ij} + \ass N^i N^j + 2\asl N^{(i}D^{j)}N + \all D^i N D^j N \,, \\
	        \cij^{ij} &= \ctr \gamma^{ij} + \css N^i N^j + 2\csl N^{(i}D^{j)}N + \cll D^i N D^j N \,.
	    \end{aligned}
	\end{align}
Here, the coefficients are functions of $N$ and $X_s$. 
The superscripts of the coefficients are assigned in a similar manner to Eq.~(\ref{C23}).

\end{itemize}

\subsubsection{Restrictions on the functions $U_0$, $U_3$, $V_0$, and $V_1$}

The restrictions on $\umetric^{\mu\nu}$ above can be translated to those on the functions $U_a$ in Section \ref{ss:ug}. 
First, from Eq.~(\ref{hatg00}), the $00$ component fixes $V_0$ in Eq.~(\ref{V0def}) as
	\begin{align}\label{V0}
		V_0 = 
        \frac{\att}{1 + \ett \rho}
        \,.
	\end{align}

Substituting this result to Eqs.~(\ref{hatg0i}), 
the solution of $\umetric^{0i}$ [Eq.~(\ref{g0i sol})] implies
	\begin{align}
        \frac{\att}{1 + \ett \rho}N^i
        + V_1 D^i N =
            \begin{cases}
		        \ati^i  \rho + \cti^i & (\ett = 0) \\
		        -\frac{\ati^i}{\ett (1 + \ett \rho)} + \cti^i & (\ett \neq 0)
		    \end{cases} 
        \,,
	\end{align}
Decomposing $\ati^i$ and $\cti^i$ as Eq.~(\ref{C23}), 
we find
    \begin{align}
	   \begin{cases}
		  \att = \as \rho + \cs\,,
		    \quad
		  V_1=\al \rho + \cl & (\ett =0) \\
		  \frac{\att}{1 + \ett \rho} = -\frac{\as}{\ett (1 + \ett \rho)} + \cs  \,, 
		    \quad
		  V_1= -\frac{\al}{\ett (1 + \ett \rho)}  + \cl &  (\ett \neq 0)
		\end{cases}\,,
	\end{align}
by comparing the coefficients. 
Since these relations are identities for $\rho$, 
the functions $\as\,,\cs$ are fixed as
    \begin{align}\label{ascs}
		\begin{cases}
		    \as=0 \,, 
    		\quad
    		\cs = \att & (\ett =0) \\
    		\as=\ett \att \,,
    		\quad
    		\cs=0  & (\ett \neq 0)
    	\end{cases}
	\end{align}
from the first equation. 
The function $V_1$ is fixed as
    \begin{align}\label{V1}
        V_1 = 
        \begin{cases}
		    \al \rho + \cl & (\ett =0) \\
		    -\frac{\al}{\ett(1 + \ett \rho)}  + \cl &  (\ett \neq 0)
        \end{cases}
    \end{align}
Substituting the above results (\ref{V0}), (\ref{ascs}), and (\ref{V1}) to Eqs.~(\ref{hatgij}) and (\ref{sol gij}), 
we find
	\begin{align}   \label{ijeq}
		U_0 \gamma^{ij} + U_3 D^i N D^j N
		 =
    		-\frac{[\al]^2 }{\att}\rho^2D^i N D^j N +  \aij^{ij} \rho + \cij^{ij} + \att N^iN^j  + 2V_1N^{(i} D^{j)}N 
    		\quad  (\ett = 0) 
	\end{align}
for the $\Gamma = 0$ case and
	\begin{align}   \label{ijeq}
		U_0 \gamma^{ij} + U_3 D^i N D^j N
		 =
    		-\frac{[\al]^2}{\att \ett^2(1 + \ett \rho)} D^i N D^j N + \aij^{ij} \rho + \cij^{ij} + 2\cl N^{(i} D^{j)}N  
      \quad (\ett \neq 0) 
	\end{align}
for the $\Gamma \neq 0$ case. 
Decomposing $\aij^{ij}$ and $\cij^{ij}$ as Eq.~(\ref{C45}) and comparing the coefficients, 
we find
    \begin{align}
        \begin{cases}
             \ass  =0 \,,  \quad \css  - \att =0 \,, \quad \asl +\al=0 \,,  \quad \csl  + \cl=0 & (\ett =0) \\
             \ass  =0 \,, \quad \css  =0\,, \quad \asl =0 \,,\quad \csl  + \cl=0 & (\ett \neq 0) 
        \end{cases}
        \,,
    \end{align}
and the functions $U_0 \,, U_3$ are fixed as
    \begin{align}
        U_0 =
        \begin{cases}
		    \atr \rho + \ctr & (\ett =0) \\
		    \atr \rho + \ctr &  (\ett \neq 0)
        \end{cases}
        \,, \qquad  
      U_3 =
        \begin{cases}
		    -\frac{[\al]^2 }{\att}\rho^2 + \all  \rho + \cll  & (\ett =0) \\
		    -\frac{[\al]^2}{\att \ett^2 (1 + \ett \rho) } + \all  \rho + \cll   &  (\ett \neq 0)
        \end{cases}
        \,.
    \end{align}

Gathering all the results in this section, we find the following restrictions on the functions $U_0, U_3, V_0, V_1$. 
The corresponding restriction on the matter metric (\ref{bargdef2}) can be read from the relations in Appendix \ref{a:FUrelations}. 
We do not give its explicit form because it is cumbersome and not insightful. 
In section~\ref{s:covariant}, 
we will give a simpler form after imposing the invertibility conditions. 

\begin{itemize}

\item $\ett=0$ case:
	\begin{align}
		U_0 &= \atr \rho + \ctr \,,
		    \label{U0_C0zero}
		\\
		U_3 &= -\frac{[\al]^2}{\att}\rho^2 + \all  \rho + \cll \,,
		    \label{U3_C0zero}
		\\
		V_0 &= 
		\att \,,
		    \label{V0_C0zero}
		\\
		V_1 &= \al \rho  + \cl\,.
		    \label{V1_C0zero}
	\end{align}
The coefficients $\att, \al, \cl, \atr, \ctr, \all, \cll$ are arbitrary functions of $N$ and $X_s$.

\item $\ett \neq 0$ case:
	\begin{align}
		U_0 &= \atr \rho + \ctr \,,
		    \label{U0_C0nonzero}
		\\
		U_3 &= 
		-\frac{[\al]^2}{\att \ett^2 (1+ \ett \rho)} + \all  \rho + \cll \,,
		    \label{U3_C0nonzero}
		\\
		V_0 &= 
		\frac{\att}{1+ \ett \rho} \,,
		    \label{V0_C0nonzero}
		\\
		V_1 &= -\frac{\al}{\ett (1 + \ett \rho)}  + \cl\,.
		    \label{V1_C0nonzero}
	\end{align}
The coefficients $\ett, \att, \al, \cl, \atr, \ctr, \all, \cll$ are arbitrary functions of $N$ and $X_s$.

\end{itemize}

\section{Invertibility conditions}
\label{s:invertible}

In this section, we examine the conditions to make the transformation $g_{\mu\nu} \to \bar{g}_{\mu\nu}$ invertible for the metric (\ref{bargdef2}). 
As explained in Section \ref{ss:action}, this corresponds to assuming the existence of the minimal coupling frame. 
The invertibility conditions have been derived in Ref.~\cite{Takahashi:2021ttd} (Eq.~(29) therein) as
    \begin{align}   \label{eq:takahashi29}
        \det(\bar{g}_{\mu\nu}) \neq 0 \,, \quad 
        F_0 \neq 0 \,, \quad \overline{X}_X \neq 0 \,, \quad 
        \overline{X}_Y = \overline{X}_Z = 0 \,, \quad 
        \left| \frac{\pd (\overline{Y},\overline{Z})}{\pd (Y,Z)} \right| \neq 0 \,,
    \end{align}
where $\overline{X}, \overline{Y}, \overline{Z}$ are defined for the new metric $\bar{g}_{\mu\nu}$ as
    \begin{align}
        \overline{X} \equiv \bar{g}^{\mu\nu}\phi_\mu \phi_\nu \,, \quad 
        \overline{Y} \equiv  \bar{g}^{\mu\nu}\phi_\mu \overline{X}_\nu \,, \quad 
        \overline{Z} \equiv \bar{g}^{\mu\nu}\overline{X}_\mu \overline{X}_\nu \,.
    \end{align}
Since the degeneracy conditions in the previous section only restrict the $\rho$-dependence, 
we first consider the invertibility conditions related to $\rho$. 
In the unitary gauge, $\bar X$ is simply given by $\bar X = \bar{g}^{00}$. 
Switching the independent variables from $X,Y,Z$ to $N,\rho,X_s$ 
we thus find that $\bar{g}^{00}$ does not depend on $\rho$ from the forth condition in Eq.~(\ref{eq:takahashi29}):
    \begin{align}\label{eq:inv cond g00}
        (\bar{g}^{00})_{,\rho} = 0 \,.
    \end{align}
Recalling that $\umetric^{\mu\nu} \equiv \sqrt{-\bar{g}}\bar{g}^{\mu\nu}$, 
this condition is expressed as a condition on $\umetric^{\mu\nu}$:
    \begin{align}   \label{g00const}
        [ \{ -\det(\umetric^{\mu\nu}) \}^{-\frac{1}{2}}\umetric^{00} ]_{,\rho} = 0 \,.
    \end{align}
With $\umetric^{00}= -V_0 \neq 0$ (Eqs.~(\ref{hatg00}) and (\ref{eq:g00req})), 
this equation can be integrated as
    \begin{align}   \label{g00const2}
        \det(\umetric^{\mu\nu}) = C_g V_0^2 \,,
    \end{align}
where $C_g$ is a function of $N$ and $X_s$. 
Using the expression of $\det(\umetric^{\mu\nu})$ in Appendix \ref{a:det}, 
we obtain the following condition for the invertibility
    \begin{align}   \label{invcond}
        N^2 U_0{}^2
        \left[ U_0 V_0 + (V_1{}^2 + V_0 U_3) X_s \right] = g C_g V_0^2 \,,
    \end{align}
with $g \equiv \det(g_{\mu\nu})$. 

\subsubsection{Restrictions on the functions $U_0$, $U_3$, $V_0$, and $V_1$}

We examine how the condition (\ref{invcond}) restricts the functions $U_0$, $U_3$, $V_0$, and $V_1$ separately for $\ett = 0$ and $\ett \neq 0$. 
The remaining conditions in Eq.~(\ref{eq:takahashi29}) will be examined in the next section. 

\begin{itemize}

\item $\ett=0$ case:

In this case, $V_0$ does not depend on $\rho$ (see Eq.~(\ref{V0_C0zero})). 
Therefore, the invertibility condition (\ref{invcond}) implies that 
    \begin{align}   \label{dethatg_C00}
        U_0{}^2 \left[ U_0 V_0 + (V_1{}^2 + V_0 U_3) X_s \right]
    \end{align}
is independent of $\rho$. 
Taking into account that $U_0, U_3, V_1$ are polynomials of $\rho$ (see Eqs.~(\ref{U0_C0zero})-(\ref{V1_C0zero})), 
we also find that 
    \begin{align}
        U_0 &= \atr \rho + \ctr \\
        V_1^2+V_0U_3 &= 
        (2\al\cl + \att \all)\rho + [\cl]^2+\att\cll
    \end{align}
should be independent of $\rho$. 
Thus,
    \begin{align}
        \atr = 0 \,, \quad 
        2\al\cl + \att \all = 0 \,.
    \end{align}
Applying these restrictions, $U_0,V_0,V_1$ are given by
    \begin{align}
    \label{eq:U0V0V1 C0=0 invertibility}
        U_0 = \ctr \,, \quad
        V_0 =\att \,, \quad
        V_1 = \al \rho + \cl \,,
    \end{align}
and $U_3$ is determined by $V_1$ as
    \begin{align}\label{eq:U3 C0=0 invertibility}
        U_3 = -\frac{V_1^2}{\att} + \cll + \frac{[\cl]^2}{\att} \equiv -\frac{V_1^2}{\att} + \cu3 \,,
    \end{align}
where we have introduced $\cu3 \equiv \cll + [\cl]^2/\att$, 
which is independent of $\rho$. 
Then, $C_g$ becomes independent of $\rho$,
    \begin{align}
        C_g =  \frac{1}{g}\frac{(N \ctr)^2}{\att} ( \ctr + \cu3 X_s ) \,,
        \label{Cg}
    \end{align}
as required. 

\item $\ett \neq 0$ case:

In this case, 
$V_0 \propto (1+ \ett \rho)^{-1}$
(see Eq.~(\ref{V0_C0nonzero})).
Therefore, the invertibility condition (\ref{invcond}) implies that 
    \begin{align}   \label{hatdetg_C0neq0}
        U_0{}^2 \left[ U_0 V_0 + (V_1{}^2 + V_0 U_3) X_s \right]
        \propto 
        (1 + \ett \rho)^{-2} \,.
    \end{align}
From Eqs.~(\ref{U0_C0nonzero})-(\ref{V1_C0nonzero}), the quantities $U_0V_0$ and $V_1^2 + V_0 U_3$ in the bracket are expressed as
    \begin{align}
        U_0 V_0 &= \frac{\att (\atr \rho + \ctr)}{1+\ett \rho} \,, \\
        V_1^2 + V_0 U_3 &= \frac{\att (\all \rho + \cll) + 2\ett^{-1}\al\cl}{1 + \ett \rho} + [\cll]^2 \,.
    \end{align}
Notably, there is no term proportional to $(1 + \ett \rho)^{-2}$ due to a cancellation. 
Since $U_0$ is the linear function of $\rho$, 
we find that the quantity on the left-hand side should identically vanish to satisfy the condition (\ref{hatdetg_C0neq0}).
This leads to
    \begin{align}
        \det(\umetric^{\mu\nu}) = 0 \quad \Rightarrow \quad \det(\bar{g}_{\mu\nu}) = 0 \,.
    \end{align}
However, this result violates the invertivility condition $\det(\bar{g}_{\mu\nu}) \neq 0$ in Eq.~(\ref{eq:takahashi29}). 
Therefore, there is no solution of the degeneracy and invertivility conditions in the $\ett \neq 0$ case. 

\end{itemize}

In summary, when both the degeneracy and invertibility conditions are imposed, 
$\umetric^{\mu\nu}$ should have the following form (see Eq.~(\ref{eq:umetric pN}) and Eqs.~(\ref{eq:U0V0V1 C0=0 invertibility})-(\ref{eq:U3 C0=0 invertibility})):
    \begin{align}   \label{eq:umetric_di2}
		\umetric^{\mu\nu} = 
    \ctr (g^{\mu\nu}+n^\mu n^\nu) 
    - \att \left(N n^\mu - \frac{V_1}{\att} \pN^\mu \right) \left(N n^\nu - \frac{V_1}{\att}\pN^\nu \right)  
    + \cu3 \pN^\mu \pN^\nu \,,
	\end{align}
with $V_1 =\al \rho + \cl$. 
Its determinant is given by
(see Eqs.~(\ref{g00const2}) and (\ref{Cg}))
    \begin{align}   \label{eq:udet_di}
        \det(\umetric^{\mu\nu}) = 
        \det(g^{\mu\nu})\att \left(N \ctr\right)^2 \left( \ctr + \cu3 X_s \right) 
        \equiv \det(g^{\mu\nu})\left(\ctr\right)^2 \detU \,.
    \end{align}
While $\umetric^{\mu\nu}$ depends on $\rho$ through $V_1$, its determinant is independent of $\rho$.

\subsubsection{Restriction on the matter metric $\bar{g}_{\mu\nu}$}

From the result (\ref{eq:umetric_di2}) and $\umetric^{\mu\nu}=\sqrt{-\bar{g}}\bar{g}^{\mu\nu}$ ($\det(\umetric^{\mu\nu})=\bar{g}$), 
we can find the form of the matter metric $\bar{g}_{\mu\nu}$ with the degenerate and invertibility conditions in the unitary gauge. 
First, the inverse matter metric $\bar{g}^{\mu\nu}$ is written as
    \begin{align} \label{eq:reduced gbarUU}
        \bar{g}^{\mu\nu} 
        &= [-\det(\umetric^{\mu\nu})]^{-\frac{1}{2}} \umetric^{\mu\nu} \nm \\
        &= \frac{\sqrt{-g}}{\detU^{\frac{1}{2}}\ctr}
        \left[  \ctr (g^{\mu\nu}+n^\mu n^\nu) 
        - \att \left(N n^\mu - \frac{V_1}{\att} \pN^\mu \right) \left(N n^\nu - \frac{V_1}{\att}\pN^\nu \right)
        + \cu3 \pN^\mu \pN^\nu\right]
        \,,
    \end{align}
by using Eqs.~(\ref{eq:umetric_di2}) and (\ref{eq:udet_di}). 
Then, the matter metric $\bar{g}_{\mu\nu}$ is computed as
    \begin{align}   \label{eq:reduced gbar}
        \bar{g}_{\mu\nu} 
        = \frac{\ctr}{\sqrt{-g}\,\detU^{\frac{1}{2}}} 
        \left[ 
            \frac{\detU}{\ctr} \gamma^{(\perp)}_{\mu\nu} 
            - \frac{\detU}{N^2 \att} n_\mu n_\nu
            + X_s \att \left(\frac{N \pN_\mu}{X_s} - \frac{V_1}{\att} n_\mu \right)\left(\frac{N \pN_\nu}{X_s} - \frac{V_1}{\att} n_\nu \right)
        \right]
        \,,
    \end{align}
where the metric $\gamma^{(\perp)}_{\mu\nu}$ represents the components of $g_{\mu\nu}$ orthogonal to $n^\mu$ and $\pN^\mu$: 
    \begin{align}
        \gamma^{(\perp)}_{\mu\nu} \equiv g_{\mu\nu} + n_\mu n_\nu - \frac{\pN_\mu \pN_\nu}{X_s} \,.
    \end{align}
It might be insightful to write down the line element for the matter metric (\ref{eq:reduced gbar}):
    \begin{align}   \label{eq:reduced gbar ds}
       \bar{g}_{\mu\nu} {\rm d}x^\mu {\rm d}x^\nu
        = -\bar{N}^2{\rm d}t^2 
        + 
        \bar{\gamma}^{(\parallel)}
            \left({\rm d}x_{\parallel} + \bar{N}^{(\parallel)} {\rm d}t \right)^2
        + (-\detU/g)^{\frac{1}{2}} \gamma^{(\perp)}_{\mu\nu}{\rm d}x^\mu {\rm d}x^\nu 
        \,,
    \end{align}
where ${\rm d}x_{\parallel} \equiv \pN_\mu {\rm d}x^\mu/\sqrt{X_s}$. 
Here, we have defined
    \begin{align}\label{eq:ds components}
        \bar{N}^2 \equiv \frac{\detU^{\frac{1}{2}}\ctr}{\sqrt{-g}\att} \,, \quad 
         \bar{\gamma}^{(\parallel)} \equiv  \frac{1}{\sqrt{-g}}\left(\frac{N \ctr}{\bar{N}}\right)^2 \,,
    \end{align}
and
    \begin{align}\label{eq:ds shift}
        \bar{N}^{(\parallel)} \equiv \frac{\sqrt{X_s}}{\att}V_1 =\frac{\sqrt{X_s}}{\att} \left(\al \rho + \cl \right) \quad \text{[Eq.~(\ref{eq:U0V0V1 C0=0 invertibility})]} \,.
    \end{align}
We see that spacetime is divided into two parts in the line element: the linear span of $\{n_\mu, \pN_\mu \}$ and its orthogonal complement. 
This reflects the fact that the disformal part (the last three terms) in the transformation (\ref{bargdef2}) is spanned by two vectors $\phi_\mu$ and $X_\mu$ (see Eq.~(\ref{nmuXmucov})). 
Then, the line element of the orthogonal complement (the last term) is only conformally transformed in Eq.~(\ref{eq:reduced gbar ds}). 
On the other hand, all components are independently transformed for ${\rm Span}\{n_\mu, \pN_\mu \}$. 
A distinctive feature is that the $\rho$ dependence only appears in the shift vector because the coefficients $\detU$, $\ctr$, $\al$, and $\cl$ are functions of $N$ and $X_s$. 
Therefore, it is easy to see that the invertible condition (\ref{eq:inv cond g00}) is satisfied. 

\section{Covariant form of the matter metric with the degenerate and invertibility conditions}
\label{s:covariant}

In this section, we write down the covariant form of the matter metric $\bar{g}_{\mu\nu}$, {\it i.e.}, the functions $F_a~(a=0,1,2,3)$ in Eq.~(\ref{bargdef2}), that satisfies both the degeneracy conditions and the invertible conditions. 

The covariant form is obtained by replacing the variables in the unitary gauge as
    \begin{align}   \label{covariantized scalars}
        N \to \frac{1}{\sqrt{-X}} \,, \quad \rho \to \frac{Y}{2(-X)^{\frac{3}{2}}} \,, \quad X_s \to \frac{Y^2-XZ}{4X^4} \,,
    \end{align}
and
    \begin{align}
        n_\mu \to -\frac{\phi_\mu}{\sqrt{-X}} \,, \quad 
        \pN_\mu \to \frac{1}{2(-X)^{\frac{3}{2}}}\left(X_\mu - \frac{Y}{X}\phi_\mu\right) \,.
        \label{nmuXmucov}
    \end{align}
Since the coefficients in Eq.~(\ref{eq:ds components}) are independent of $\rho$, 
all the covariantized coefficients should be functions of
only $\phi$, $X$, and the specific combination $W \equiv Z-Y^2/X$ (see Eq.~(\ref{covariantized scalars})).
Therefore, 
the covariant form of the matter metric $\bar{g}_{\mu\nu}$ can be expressed as%
\footnote{We extracted the factor $XW=XZ-Y^2$ in the last term to simplify the latter expressions (\ref{eq:resultant restrictions extended}).}
    \begin{align}   \label{eq:resultant transformation extended}
	    \bar g_{\mu\nu}=F_\phi^{\rm U}(\phi ,X, W)\phi_\mu\phi_\nu 
     + F_{\parallel}^{\rm U}(\phi ,X,W) \left( X_\mu - V^{\rm U} \phi_\mu \right) \left( X_\nu - V^{\rm U} \phi_\nu \right)
     + XW F^{\rm U}_{\perp}(\phi,X,W)\gamma^{(\perp)}_{\mu\nu} 
	\,,
    \end{align}
where $V^{\rm U}=V^{\rm U}(\phi, X, Y, W)$ is a linear function in $Y$: 
    \begin{align}
        V^{\rm U}(\phi, X, Y, W) =  A^{\rm U}(\phi, X, W) Y +  C^{\rm U}(\phi, X, W) \,.
    \end{align}
Here, the metric $\gamma^{(\perp)}_{\mu\nu}$ denotes the components orthogonal to $\phi_\mu$ and $X_\mu$. 
We can explicitly write it as
    \begin{align}
        \gamma^{(\perp)}_{\mu\nu} = g_{\mu\nu} - \frac{\phi_\mu\phi_\nu}{X} - \frac{ \pX_\mu \pX_\nu}{W} \,,
    \end{align}
in terms of $\phi_\mu$ and
    \begin{align} \label{def:pX}
        \pX_\mu \equiv X_\mu - \frac{Y}{X}\phi_\mu \quad  ( \pX_\mu \pX^\mu = W ) \,, 
    \end{align}
which corresponds to the components of $X_\mu$ orthogonal to $\phi_\mu$. 
The functions $F_a^{\rm U}(\phi,X,W)~(a=\phi, \parallel, \perp)$ and $V^{\rm U}(\phi, X, Y, W)$ can be chosen independently. 
Comparing the metric (\ref{eq:resultant transformation extended}) with Eq.~(\ref{bargdef2}), 
we find that the functions $F_a(\phi,X,Y,Z)$ are given by
    \begin{align}\label{eq:resultant restrictions extended}
        \begin{aligned}
            &F_0(\phi,X,Y,Z)= XW F_\perp^{\rm U}(\phi, X,W)
            \,,\ \ 
            F_1(\phi,X,Y,Z)=F_\phi^{\rm U}(\phi ,X,W) + [V^{\rm U}]^2 F_\parallel^{\rm U}(\phi ,X,W) - Z F_\perp^{\rm U}(\phi,X,W)
            \,,\\
            &F_2(\phi,X,Y,Z)=-V^{\rm U} F_\parallel^{\rm U}(\phi, X, W) + Y F_\perp^{\rm U}(\phi,X,W)
            \,,\ \ 
            F_3(\phi,X,Y,Z)=F_\parallel^{\rm U}(\phi, X,W) - X F_\perp^{\rm U}(\phi,X,W)
            \,.
        \end{aligned}
    \end{align}
The dependence on the higher-derivative terms $Y$ and $Z$ 
in the functions $F_a~(a=0,1,2,3)$ is restricted as the relations above: the functions $F_a$ cannot be freely chosen.  
For example, $F_1=0$ ($F_2=0$) leads to $F_2=F_3=0$ ($F_3=0$) because the relations above require $F_\phi^{\rm U}=F_\parallel^{\rm U}=F_\perp^{\rm U}=0$ ($F_\parallel^{\rm U}=F_\perp^{\rm U}=0$) in this case.

We examine the remaining invertibility conditions in Eq.~(\ref{eq:takahashi29}). 
The conditions $\overline{X}_Y=\overline{X}_Z=0$ imply that $\overline{X} = \bar{g}^{\mu\nu}\phi_\mu \phi_\nu$ is independent of $W$ as well as $Y$.  
Substituting the restricted form of the matter metric (\ref{eq:resultant transformation extended}), 
we find $\overline{X} = [F_\phi^{\rm U}]^{-1}$. 
Thus, the invertibility conditions $\overline{X}_Y=\overline{X}_Z=0$ are satisfied when
    \begin{align}
        F_\phi^{\rm U} = F_\phi^{\rm U}(\phi, X) \,.
    \end{align}
The other conditions state that some quantities do not vanish. 
Therefore, they are satisfied except for specific forms of the functions $F_\phi^{\rm U}$, $F_\parallel^{\rm U}$, $F^{\rm U}_\perp$, and $V^{\rm U}$. 
Since the explicit forms are cumbersome, we do not show them for a general case. 
We will closely examine them only for a special case below. 

Finally, we would like to remark that our result contains that in Ref.~\cite{Takahashi:2022mew},\footnote{We thank K. Takahashi, M. Minamitsuji, and H. Motohashi for sharing information on their project before submitting the first version of our and their papers to arXiv. }
which is obtained by requiring that the matter metric $\bar{g}_{\mu\nu}$ is independent of $\dot{N}$ in the unitary gauge:  
when the function $V^{\rm U}$ is chosen to satisfy
    \begin{align}
        X_\mu - V^{\rm U}\phi_\mu = \pX_\mu - C^{\rm U} \phi_\mu \quad (
        \Leftrightarrow~
        A^{\rm U}=1/X)\,,
    \end{align}
the metric (\ref{eq:resultant restrictions extended}) can be reorganized as
    \begin{align}   \label{eq:resultant transformation}
	    \bar g_{\mu\nu}=F^{\rm U}_0(\phi,X,W)g_{\mu\nu}+F_1^{\rm U}(\phi ,X, W)\phi_\mu\phi_\nu +2F^{\rm U}_2(\phi ,X, W)\phi_{(\mu}\pX_{\nu )}+F_3^{\rm U}(\phi ,X,W)\pX_\mu\pX_\nu
	\,.
     \end{align}
In contrast that three scalars $X, Y, Z$ can be constructed from $\phi_\mu$ and $X_\mu$, 
we can only construct two scalars $X, W$ from  $\phi_\mu$ and $\pX_\mu$ because $\phi_\mu \pX^\mu = 0$. 
Therefore, 
this special class is easier to handle because $Y$ never appears in any manipulations ({\it e.g.,} matrix product, functional composition) and only $\phi, X, W$ appears as independent variables. 
It is also easy to see that the second derivative along the time direction does not appear in this special class. 
We can rewrite $\pX_\mu$ as
    \begin{align}
        \pX_\mu = 2\phi^\alpha \left(\delta_\mu{}^\beta -\frac{\phi_\mu \phi^\beta}{X}\right)\nabla_\alpha\nabla_\beta\phi  \,.
    \end{align}
This expression implies that $\pX_\mu$ is proportional to $\nabla_\alpha\nabla_\beta\phi$ projected perpendicular to the direction $\phi_\alpha \equiv \nabla_\alpha\phi$.
Therefore, in the unitary gauge $\phi=t$, the second derivative along
the time direction does not appear as stated above.

Before closing this section, 
let us examine the remaining invertibility conditions in Eq.~(\ref{eq:takahashi29}) for the special class (\ref{eq:resultant transformation}). 
For the restricted form (\ref{eq:resultant transformation}), 
the conditions (\ref{eq:takahashi29}) are reduced to
    \begin{align}   \label{eq:takahashi29_restrict}
        \det(\bar{g}_{\mu\nu}) \neq 0 \,, \quad F_0^{\rm U} \neq 0 \,, \quad \overline{X}_X \neq 0 \,, \quad \overline{X}_W = 0 \,, \quad \overline{W}_W \neq 0 \,.
    \end{align}
From the first two conditions, we obtain the following conditions on $F_a^{\rm U}$:
    \begin{align}
        {\cal F}^{\rm U} \equiv \frac{\det(\bar{g}_{\mu\nu})}{[F_0^{\rm U}]^2\det(g_{\mu\nu})} = [F_0^{\rm U}]^2 + F_0^{\rm U}(XF_1^{\rm U}+WF_3^{\rm U}) - XW([F_2^{\rm U}]^2-F_1^{\rm U}F_3^{\rm U}) \neq 0 \,.
    \end{align}
In writing down the remaining conditions, 
it is convenient to introduce the functions $f_a^{\rm U}~(a=0,1,2,3)$ for the inverse metric through
    \begin{align}
        \bar{g}^{\mu\nu} = f^{\rm U}_0(\phi,X,W)g^{\mu\nu}+f_1^{\rm U}(\phi ,X, W)\phi^\mu\phi^\nu +2f^{\rm U}_2(\phi ,X, W)\phi^{(\mu}\pX^{\nu )}+f_3^{\rm U}(\phi ,X,W)\pX^\mu\pX^\nu
        \,.
    \end{align}
The functions $f_a^{\rm U}$ are expressed in terms of $F_a^{\rm U}$ as
    \begin{align}
        \begin{aligned}
            &f_0^{\rm U} = \frac{1}{F_0^{\rm U}} \,, \quad f_1^{\rm U} = -\frac{F_0^{\rm U}F_1^{\rm U} - W([F_2^{\rm U}]^2-F_1^{\rm U}F_3^{\rm U})}{ F_0^{\rm U} {\cal F}^{\rm U} } \,, \\
            &f_2^{\rm U} = -\frac{F_2^{\rm U}}{{\cal F}^{\rm U} } \,, \quad f_3^{\rm U} = -\frac{F_0^{\rm U}F_3^{\rm U} - X([F_2^{\rm U}]^2-F_1^{\rm U}F_3^{\rm U})}{ F_0^{\rm U} {\cal F}^{\rm U} } \,.
        \end{aligned}
    \end{align}
Using $f_a^{\rm U}$, we can express $\overline{X}$ as
    \begin{align}
        \overline{X} = X(f_0^{\rm U} + X f_1^{\rm U}) = \frac{X(F_0^{\rm U} + W F_3^{\rm U})}{{\cal F}^{\rm U}}\,.
    \end{align}
The third and forth conditions in Eq.~(\ref{eq:takahashi29_restrict}) imply that the combination above depends on $X$ and not on $W$, respectively. 

Because $\overline{X}=\overline{X}(\phi,X)$, $\overline{X}_\mu$ can be computed as $\overline{X}_\mu = \overline{X}_\phi \phi_\mu + \overline{X}_X X_\mu$. 
Here, the subscripts $\phi, X$ indicate the derivatives with respect to the corresponding variables. 
After some manipulation, we find
    \begin{align}
        \overline{\pX}_\mu \equiv \overline{X}_\mu - (\overline{Y}/\overline{X})\phi_\mu
        = \left[ -f_2^{\rm U} \frac{XW}{\overline{X}} \phi_\mu + \pX_\mu \right]\overline{X}_X \,,
    \end{align}
and $\overline{W}$ can be expressed as
    \begin{align}
            \overline{W} 
            = \left\{ f_0^{\rm U} + W \frac{f_0^{\rm U}f_3^{\rm U} -X([f_2^{\rm U}]^2-f_1^{\rm U}f_3^{\rm U})}{f_0^{\rm U} + f_1^{\rm U}X}\right\}\overline{X}_X W
            = \frac{\overline{X}_X W}{F_0^{\rm U} + W F_3^{\rm U}} \,.
    \end{align}
The last condition in Eq.~(\ref{eq:takahashi29_restrict}) implies that the combination above depends on $W$. 
Note that $Y$ does not appear in the mapping $(X,W) \leftrightarrow (\overline{X}, \overline{W})$. 

As an illustration, let us consider the case with $F_0^{\rm U}=1$, $F_1^{\rm U}=0$, $F_2^{\rm U}=0$ and $F_3^{\rm U} \neq 0$ as a counterpart of the example in Ref.~\cite{Takahashi:2021ttd} (Section IIIA therein). 
In this case, $\overline{X}$ and $\overline{W}$ become
    \begin{align}
        \overline{X} = X \,, \quad \overline{W} = \frac{W}{1+WF_3^{\rm U}} \,.
    \end{align}
Thus, the invertibility conditions (\ref{eq:takahashi29_restrict}) are solved by
    \begin{align}
        \bar{g}_{\mu\nu} = g_{\mu\nu} + F_3^{\rm U}(\phi,X,W)\pX_\mu\pX_\nu \,,
    \end{align}
for any function $F_3^{\rm U}(\phi,X,W)$ with $F_3^{\rm U} \neq P(\phi,X)-W^{-1}$ for some function $P(\phi,X)$. 
We can also write it in the form (\ref{bargdef2}) as
    \begin{align}
		\bar{g}_{\mu\nu} = g_{\mu\nu} + F_1(\phi, X,Y,Z) \phi_\mu \phi_\nu + 2F_2(\phi, X,Y,Z) \phi_{(\mu}X_{\nu)} + F_3(\phi, X,Y,Z) X_\mu X_\nu~,
	\end{align}
with 
\begin{equation}
F_1=(Y/X)^2 F_3^{\rm U}(\phi,X,Z-Y^2/X)~, 
\quad
F_2=-(Y/X) F_3^{\rm U}(\phi,X,Z-Y^2/X)~,
\quad
F_3= F_3^{\rm U}(\phi,X,Z-Y^2/X)~.
\end{equation}
As we have stated, the functions $F_a~(a=1,2,3)$ are related with each other to realize the cancellation between the higher-derivative terms. 

\section{Summary}
\label{s:summary}

In this work, we examined how a matter coupling changes the degeneracy for new theories of modified gravity generated by the metric transformation (\ref{bargdef}), which is a generalization of the disformally-transformed metric.  
Even when the Ostrogradsky mode does not exist in a theory only with the gravity action, 
it may revive once the matter coupling is introduced unless the matter coupling satisfies certain conditions.
We derived such conditions to avoid re-appearance of the Ostrogradsky mode due to the matter coupling.

To derive the conditions to avoid the Ostrogradsky mode, 
we started from the Horndeski action with a matter coupling (\ref{Lmatter}), 
in which a free scalar field $\sigma$ is coupled to a metric $\bar{g}_{\mu\nu}$ defined by Eq.~(\ref{bargdef}). 
When the metric transformation (\ref{bargdef}) from $g_{\mu\nu}$ to $\bar{g}_{\mu\nu}$ is invertible, 
this theory is equivalent to a new gravity theory which is generated from the Horndeski theory through the metric transformation (\ref{bargdef}) and has a minimal coupling to the matter scalar field $\sigma$. 
Thus, the action (\ref{Lag}) incorporates the DHOST class I theories and their generalizations proposed in Ref.~\cite{Takahashi:2021ttd} with a minimally coupled matter scalar field. 
Working in the unitary gauge, we found that the degeneracy conditions of this system are exactly solvable and the matter metric $\bar{g}_{\mu\nu}$ is restricted. 
We would like to note that this result would be nontrivial when we started from the action in the new gravity frame ({\it i.e.,} the minimal coupling frame). 
In this frame, the matter action does not contain the higher derivatives nor affect the degeneracy of the kinetic matrix.
To find the same restrictions, we will need to perform the full constraint analysis (see Appendix \ref{a:frames}). 

We derived the conditions on the matter metric without and with assuming the invertibility in section~\ref{s:dc} and in section~\ref{s:invertible}, respectively. 
As explained in section~\ref{s:covariant}, 
the second-derivative terms of the gravity scalar field $\phi$ in the matter metric has a certain structure when both the degeneracy and invertibility conditions are imposed (see Eq.~(\ref{eq:resultant transformation})): 
the second derivative $\nabla_\mu \nabla_\nu \phi$ only appears with the projection onto the orthogonal complement of $\nabla_\mu \phi$.
This manifestly shows the absence of the higher time-derivative of $\phi$ in the unitary gauge $\phi=t$. 
This structure is similar to the so-called U-DHOST theories, which are degenerate in the unitary gauge but not in their covariant version \cite{DeFelice:2018ewo}. 
Therefore, the resultant theory (\ref{Lag}) with the matter metric (\ref{eq:resultant transformation}) will have an instantaneous mode in the presence of the matter field. 
The instantaneous mode does not induce an instability as the Ostrogradsky mode but will lead to a peculiar phenomenology. 
It will be interesting to investigate how the phenomenology of the theory (\ref{Lag}) is affected by the presence of the matter field in the environment. 

Some directions of further studies are in order. 
First, as a solvable problem, 
we investigated the degeneracy conditions in the presence of a free scalar field as the matter. 
Further studies are necessary to see how much the matter metric is restricted for the realistic matter fields. 
Second, 
we limited our analysis within the unitary gauge. 
If we desire to eliminate the instantaneous mode, 
the degeneracy conditions should be imposed to the covariant version of the theory. 
It will be interesting to study how much the matter metric is further restricted in this case. 
Finally, we worked only on the metric transformation given by Eq.~(\ref{bargdef}). 
In principal, 
this transformation can be generalized to incorporate more higher-derivative terms as done by Ref.~\cite{Takahashi:2021ttd}. 
Such an extension would be important to clarify the parameter space of the physically sensible theories without Ostrgradsky modes.

\section*{Acknowledgments}
We thank Ryo Namba for useful comments. 
This work is supported in part by JSPS Grant-in-Aid for Scientific Research No.\ 20H05852 (A.~N.), No.\ 17K14286, 20H05860 (R.~S.), No.\ JP18K03623, 21H05189 (N.~T.), No.\ 22K03627 (D.~Y.), No.\ 19H01891 (A.~N., R.~S., and D.~Y.). 

\appendix

\section{Explicit expressions for the metrics $\bar{g}_{\mu\nu}$ and $\umetric^{\mu\nu}$}
\label{a:FUrelations}

\subsection{Matter metric $\bar{g}_{\mu\nu}$}

We summarize some formulae on the matter metric (\ref{bargdef2}), which were given in Ref.~\cite{Takahashi:2021ttd}.

\begin{itemize}
    \item[--] Definition
        \begin{align}
             \bar{g}_{\mu\nu} = F_0 g_{\mu\nu} + F_1 \phi_\mu\phi_\nu + 2F_2 \phi_{(\mu}X_{\nu)} + F_3 X_\mu X_\nu \,.
        \end{align}
    \item[--] Determinant
        \begin{align}
            \sqrt{-\bar{g}} =F_0{\cal F}^{1/2} \sqrt{-g} \,,
        \end{align}
    where
        \begin{align}
            {\cal F} \equiv F_0^2+F_0(XF_1+2YF_2+ZF_3)+(F_2^2-F_1F_3)(Y^2-XZ) \,.
        \end{align}
    \item[--] Inverse metric
        \begin{align}
            \bar{g}^{\mu\nu} = f_0 g^{\mu\nu} + f_1 \phi^\mu\phi^\nu + 2f_2 \phi^{(\mu}X^{\nu)} + f_3 X^\mu X^\nu \,,
        \end{align}
    where
        \begin{align}
            \begin{aligned}
                &f_0 \equiv \frac{1}{F_0} \,, \quad
                f_1 \equiv
                -
                \frac{F_0F_1-Z(F_2^2-F_1F_3)}{F_0{\cal F}} \,,\\
                &f_2 \equiv
                -
                \frac{F_0F_2+Y(F_2^2-F_1F_3)}{F_0{\cal F}} \,, \quad f_3 =
                -
                \frac{F_0F_3-X(F_2^2-F_1F_3)}{F_0{\cal F}} \,.
            \end{aligned}
        \end{align}
\end{itemize}

\subsection{$\umetric^{\mu\nu} \equiv \sqrt{-\bar{g}}\bar{g}^{\mu\nu}$}

We show the relationship between the functions $U_a~(a=0,1,2,3)$ in $\umetric^{\mu\nu}$ (Eq.~(\ref{eq:umetric def})) and the functions $F_a~(a=0,1,2,3)$ in  the matter metric $\bar{g}_{\mu\nu}$ (Eq.~(\ref{bargdef2})).

\begin{itemize}
    \item[--] Expression in the unitary gauge
        \begin{align} \label{hatg_appendix}
            \umetric^{\mu\nu} = U_0 g^{\mu\nu} + U_1n^\mu n^\nu + 2U_2 n^{(\mu}\nabla^{\nu)}N + U_3 \nabla^\mu N \nabla^\nu N \,,
        \end{align}
    where
        \begin{align}
            U_0 &= \sqrt{-g}{\cal F}^{1/2} \,, \\
            U_1 &= -\frac{U_0}{N^2} \left[ \frac{F_0F_1-Z(F_2^2-F_1F_3)}{{\cal F}} \right] \,, \\
            U_2 &= 
            \hphantom{-}
            \frac{2U_0}{N^4} \left[ \frac{F_0F_2+Y(F_2^2-F_1F_3)}{{\cal F}} \right] \,, \\
            U_3 &= -\frac{4U_0}{N^6}\left[ \frac{F_0F_3-X(F_2^2-F_1F_3)}{{\cal F}} \right] \,.
        \label{tmplabel1}
        \end{align}
    \item[--] Expression of the components
        \begin{align}
		    \umetric^{00} &= -V_0 \,, \\
		    \umetric^{0i} &= V_0 N^i + V_1 D^i N \,, \\
		    \umetric^{ij} &= U_0 \gamma^{ij} - V_0 N^iN^j - 2V_1N^{(i} D^{j)}N + U_3 D^i N D^j N \,,
	    \end{align}
	where
	   \begin{align}
		    V_0 &\equiv \frac{U_0 - U_1 -2U_2N\rho - U_3 (N\rho)^2}{N^2}
		        = 
          -\frac{F_0 U_0}{{\cal F}}\left[ XF_0 
          - (Y^2-XZ)F_3 \right]
          \,, \\
		    V_1 &\equiv \frac{U_2+U_3 N\rho}{N} 
		    =
      -
      \frac{2(-X)^{3/2}F_0 U_0}{{\cal F}}(XF_2 
      +Y
      F_3)
      \,.
	\end{align}
\end{itemize}

\section{Determinant $\det(\umetric^{\mu\nu})$}
\label{a:det}

In this appendix, 
we derive the expression of $\det(\umetric^{\mu\nu})$. 
The derivation is based on the matrix determinant lemma
    \begin{align}
        \det(A^{\mu\nu} + u^{\mu}v^{\nu}) = (1+v^{\nu}(A^{-1})_{\nu\mu}u^{\mu})\det(A^{\mu\nu}) \,.
    \end{align}
For the derivation, 
we define the following tensors:
    \begin{align}
        \umetric^{\mu\nu}_{(0)}
        &=
        U_0 g^{\mu\nu}
        \label{hatg0def}
        \\
        \umetric^{\mu\nu}_{(1/2)}
        &=
        U_0 g^{\mu\nu}+ U_2 n^{\mu}\nabla^{\nu}N
        \label{g12}
        \\
        \umetric^{\mu\nu}_{(1)}
        &=
        U_0 g^{\mu\nu} + 2U_2 n^{(\mu}\nabla^{\nu)}N 
        \\
        \umetric^{\mu\nu}_{(2)}
        &=
        U_0 g^{\mu\nu} + U_1n^\mu n^\nu + 2U_2 n^{(\mu}\nabla^{\nu)}N \,.
        \label{hatg2def}
    \end{align}

Then, applying the matrix determinant lemma repeatedly to the metric (\ref{eq:umetric}), 
we find
    \begin{align}
            \frac{\det(\umetric^{\mu\nu}) }{\det(g^{\mu\nu})}
            &=
            U_0{}^4
            \left(1+ \frac{U_2}{U_0} g_{\mu\nu}\left(\nabla^\mu N\right) n^{\nu}\right)
            \left(1+ U_2 \umetric^{(1/2)}_{\mu\nu} n^{\mu} \nabla^\nu N \right)
            \left(1+ U_1 \umetric^{(1)}_{\mu\nu} n^\mu n^\nu \right)
            \left(1+ U_3 \umetric^{(2)}_{\mu\nu}\nabla^\mu N\nabla^\nu N \right)
            \nm\\
            &=
            U_0{}^2\left\{
            U_0\left( 
            U_0 
            + U_1 n^2
            + 2U_2 n\cdot\nabla N 
            + U_3 (\nabla N)^2
            \right)
            + 
            \left( U_2{}^2 - U_1 U_3\right) 
            \left[ (n\cdot \nabla N)^2 - n^2 (\nabla N)^2\right]
            \right\}
            \nm\\
            &=
            U_0{}^2\left\{
            U_0 \left[
            U_0 - U_1 - 2U_2 N \rho + U_3 \left( -N^2 \rho^2 + X_s \right)
            \right]
            + \left(U_2{}^2 - U_1 U_3\right) X_s
            \right\}
            \nm\\
            &=
            N^2 U_0{}^2\left[
            U_0 V_0 + \left( V_1{}^2 + V_0 U_3\right)X_s
            \right]
            \equiv
            U_0{}^2\detU
            \,,
            \label{eq:calU def}
    \end{align}
where
$n^2\equiv g^{\mu\nu}n_\mu n_\nu$, 
$(\nabla N)^2 \equiv g^{\mu\nu}\nabla_\mu N \nabla_\nu N$,
$n\cdot \nabla N \equiv g^{\mu\nu}n_\mu \nabla_\nu N$. 
To show the third and fourth lines, 
we used
    \begin{align}
        n^2=-1 \,, \quad
        n\cdot \nabla N = -N\rho \,,
        \quad
        (\nabla N)^2
        =
        X_s - N^2 \rho^2 \,,
        \quad
         (n\cdot \nabla N)^2 - n^2 (\nabla N)^2
         = X_s \,,
    \end{align}
and Eqs.~(\ref{V0def}) and (\ref{V1def}).
In addition, 
$\umetric^{(n)}_{\mu\nu}$ are the inverse matrices of $\umetric^{\mu\nu}_{(n)}$ in Eqs.~(\ref{hatg0def})--(\ref{hatg2def}): 
    \begin{align}
        \umetric^{(1/2)}_{\mu\nu}
        &=
        \frac{1}{U_0}\left(
        g_{\mu\nu} - \frac{U_2 n\cdot\nabla N}{U_0 + U_2 n\cdot \nabla N}
        \right)
        \label{g12inverse}
        \\
        \umetric_{\mu\nu}
        &=
        \frac{1}{U_0} g_{\mu\nu}
        +
        \frac{1}{U_0\, \mathcal{U}}
        \biggl\{
        \left[
        \left(U_2{}^2-U_1 U_3 \right) (\nabla N)^2 - U_0 U_1
        \right]n_\mu n_\nu
        \notag \\
        &\hspace{25mm}
        + 
        \left[
        \left(U_2{}^2 -U_1 U_3\right)n^2
        - U_0 U_3
        \right]
        \nabla_\mu N \nabla_\nu N
        - 2\left[
        U_0 U_2 +\left(U_2{}^2 - U_1 U_3\right) n\cdot \nabla N
        \right]
        n_{(\mu}\nabla_{\nu)}N
        \biggr\}\,,
        \label{hatginv}
    \end{align}
where $\umetric^{(n)}_{\mu\nu}$ for $n=0,1,2$ are given by setting $U_{m>n}$ to zero in Eq.~(\ref{hatginv}) ({\it e.g.,} $\umetric^{(2)}_{\mu\nu}$ is given by Eq.~(\ref{hatginv}) with $U_3=0$).
The inverse matrices $\umetric^{(n)}_{\mu\nu}$ are derived by using
the Sherman-Morrison formula
    \begin{align}
        \left[
        \left(
        A^{\mu\nu} + u^\mu v^\nu
        \right)^{-1}
        \right]_{\mu\nu}
        =
        \left(A^{-1}\right)_{\mu\nu}
        -\frac{\left(A^{-1}\right)_{\mu\alpha}u^\alpha v^\beta\left(A^{-1}\right)_{\beta\nu}}{1+v^\gamma\left(A^{-1}\right)_{\gamma\delta}u^\delta}\,.
    \end{align}
The equation (\ref{eq:reduced gbar}) can be obtained from Eq.~(\ref{eq:reduced gbarUU}) by comparing Eqs.~(\ref{hatg_appendix}) and (\ref{hatginv}) and reorganizing the expression in terms of $\pN^\mu \equiv \nabla^\mu N - N\rho\, n^\mu$.
    
\section{Structure of degeneracy in different frames}
\label{a:frames}

In the main text, we analyzed the degeneracy conditions for the action in the Horndeski frame (\ref{Lag}) (Eq.~(\ref{caseC2})). 
When we instead use the action in the new gravity frame (Eq.~(\ref{caseC1})), 
the gravity action depends on $\ddot{N}$ as well as $\dot{N}$ and we need constraints to remove them. 
In this Appendix, 
we see how the matter coupling affects the structure of degeneracy in the two different frames by using a simple example of a mechanical system:
    \begin{align}   \label{barngssytem}
            \bar{S}[N,g,\sigma] = \int{\rm d}t~\left[\bar{L}^{\rm G}(N,\dot{N},\ddot{N},g,\dot{g}) + g\dot{\sigma}^2\right] \,,
    \end{align}
where the variables $N$, $g$, $\sigma$ are not fields but the functions of $t$. 
We assume that the higher-derivative Lagrangian $\bar{L}^{\rm G}(N,\dot{N},\ddot{N},g,\dot{g})$ is generated from the Lagrangian $L^{\rm G}(N,g,\dot{g})$ through the invertible transformation $g \to \bar{g}$:
    \begin{align}\label{barLL}
        \bar{L}^{\rm G}(N,\dot{N},\ddot{N},g,\dot{g}) = L^{\rm G}(N,\bar{g},\dot{\bar{g}}) \,;~ \bar{g} = \bar{g}(g,N,\dot{N}) \,.
    \end{align}
Therefore, the action (\ref{barngssytem}) is equivalent to
        \begin{align}   \label{ngssytem}
            S[N,\bar{g},\sigma] = \int{\rm d}t~\left[L^{\rm G}(N,\bar{g},\dot{\bar{g}}) + g\dot{\sigma}^2\right] \,;~ g = g(\bar{g},N,\dot{N}) \,,
    \end{align}
where the relation $g=g(\bar{g},N,\dot{N})$ is obtained by inverting the transformation $g \to \bar{g}$. 
The action (\ref{barngssytem}) is a special class of the coupled mechanical systems analyzed in Refs.~\cite{Motohashi:2016ftl, Klein:2016aiq}.
Following the analysis in Refs.~\cite{Motohashi:2016ftl, Klein:2016aiq}, 
we show how the matter coupling $g\dot{\sigma}^2$ makes us fail to remove the derivative terms of $N$ from the equations of motion in the two frames (\ref{barngssytem}) and (\ref{ngssytem}). 

The action (\ref{ngssytem}) corresponds to the Horndeski frame (\ref{Lag}) (Eq.~(\ref{caseC2})). 
In this frame, the equations of motion are written as
    \begin{align}
        E^N &\equiv -\dif{}{t}(g_{\dot{N}}\dot{\sigma}^2) + L^{\rm G}_{N} + g_N \dot{\sigma}^2
        = -(g_{\dot{N}\dot{N}}\dot{\sigma}^2)\ddot{N} + \cdots \,, \\
        E^{\bar{g}} &\equiv  -\dif{}{t}(L^{\rm G}_{\dot{\bar{g}}}) + L^{\rm G}_{\bar{g}} + g_{\bar{g}}\dot{\sigma}^2 
        = -L^{\rm G}_{N\dot{\bar{g}}}\dot{N} + g_{\bar{g}}\dot{\sigma}^2  + \cdots \,, \\
        E^{\sigma}/2 &\equiv -\dif{}{t}(g\dot{\sigma}) 
        = -g_{\dot{N}}\dot{\sigma}\ddot{N} +\cdots \,,
    \end{align}
where only the highest-derivative term of $N$ is shown in the r.h.s of each equation of motion. 
The subscripts represent the derivatives with respect to the corresponding variables. 
It is apparent that the matter coupling affects the degeneracy conditions to remove the highest derivative $\ddot{N}$ in the equations of motion. 
In the Hamiltonian analysis, 
this corresponds to the fact that the matter coupling affects the primary constraints.

The action (\ref{barngssytem}) corresponds to the new gravity frame (Eq.~(\ref{caseC1})). 
In this frame, the equations of motion are written as
    \begin{align}
        \bar{E}^N &\equiv \hdif{2}{}{t}(\bar{L}^{\rm G}_{\ddot{N}}) - \dif{}{t}(\bar{L}^{\rm G}_{\dot{N}}) + \bar{L}^{\rm G}_{N} = \bar{L}^{\rm G}_{\ddot{N}\ddot{N}}\ddddot{N} + \cdots \,, \\
        \bar{E}^g &\equiv  -\dif{}{t}(\bar{L}^{\rm G}_{\dot{g}}) + \bar{L}^{\rm G}_{g} + \dot{\sigma}^2 = -\bar{L}^{\rm G}_{\dot{g}\ddot{N}}\dddot{N} + \cdots \,, \\
        \bar{E}^{\sigma}/2 &\equiv -\dif{}{t}(g\dot{\sigma}) = -g_{\dot{N}}\dot{\sigma}^2 \ddot{N} + \cdots \,,
    \end{align}
where only the highest-derivative term of $N$ is shown in the r.h.s of each equation of motion. 
In this frame, the highest derivative in the equations of motion is $\ddddot{N}$. 
We can show that the matter coupling does not affect the degeneracy condition for the highest derivative $\ddddot{N}$. 
When the Lagrangian $\bar{L}^{\rm G}$ is generated through the transformation as Eq.~(\ref{barLL}), 
$\bar{L}^{\rm G}_{\ddot{N}}$ and $\bar{L}^{\rm G}_{\dot{g}}$ can be written in terms of $L^{\rm G}$: 
    \begin{align}   \label{bar_to_unbar}
        \bar{L}^{\rm G}_{\ddot{N}} = L^{\rm G}_{\dot{\bar{g}}}\bar{g}_{\dot{N}} \,, \quad 
        \bar{L}^{\rm G}_{\dot{g}} = L^{\rm G}_{\dot{\bar{g}}}\bar{g}_{g} \,.
    \end{align}
Thus, the Lagrangian $\bar{L}^{\rm G}$ satisfies the condition
    \begin{align}   \label{primary_condition}
        \bar{L}^{\rm G}_{\ddot{N}} - v \bar{L}^{\rm G}_{\dot{g}} = 0 \,; \quad v \equiv \frac{\bar{g}_{\dot{N}}}{\bar{g}_g} = v(g,N, \dot{N})  \,,
    \end{align}
which also implies that the kinetic matrix
    \begin{align}
        \left(
        \begin{matrix}
            \bar{L}^{\rm G}_{\ddot{N}\ddot{N}} & \bar{L}^{\rm G}_{\dot{g}\ddot{N}}  \\
            \bar{L}^{\rm G}_{\ddot{N}\dot{g}} & \bar{L}^{\rm G}_{\dot{g}\dot{g}} 
        \end{matrix}
        \right)
    \end{align}
is degenerate. 
The condition (\ref{primary_condition}) ensures that the highest derivative $\ddddot{N}$ can be removed by taking the following combination,
    \begin{align}
        \tilde{\bar{E}}^N \equiv \bar{E}^N + \dif{}{t}(v \bar{E}^g) 
        = - \dif{}{t}\left[ \bar{L}^{\rm G}_{\dot{N}} - \dot{v}\bar{L}^{\rm G}_{\dot{g}} - v\bar{L}^{\rm G}_g - v\dot{\sigma}^2 \right] + \bar{L}^{\rm G}_{N}
        = -(\bar{L}^{\rm G}_{\dot{N}} - \dot{v}\bar{L}^{\rm G}_{\dot{g}} - v\bar{L}^{\rm G}_g)_{\ddot{N}}\dddot{N} + \cdots \,.
    \end{align}
Therefore, 
the matter coupling does not affect the degeneracy condition to remove the highest derivative $\ddddot{N}$, {\it i.e.,} the condition (\ref{primary_condition}). 
However, the matter coupling now appears in the terms with $\ddot{N}$ and $\dot{N}$ through the term $(v\dot{\sigma}^2)^{\cdot}$. 
The matter coupling appears in the degeneracy conditions for the lower derivatives. 
In the Hamiltonian analysis, 
this corresponds to the fact that the matter coupling does not affect the primary constraints but does the higher-order  (secondary, tertiary,...) constraints 
(see Ref.~\cite{Klein:2016aiq} and its reference for the relation of the Lagrangian analysis here to the Hamiltonian analysis).

\input{higherd_arXiv2.bbl}

\end{document}

%% file: higherd_arXiv2.bbl
\providecommand{\noopsort}[1]{}\providecommand{\singleletter}[1]{#1}%
%

%% file: higherd_arXiv2.bbl
\begin{thebibliography}{34}%
\makeatletter
\providecommand \@ifxundefined [1]{%
 \@ifx{#1\undefined}
}%
\providecommand \@ifnum [1]{%
 \ifnum #1\expandafter \@firstoftwo
 \else \expandafter \@secondoftwo
 \fi
}%
\providecommand \@ifx [1]{%
 \ifx #1\expandafter \@firstoftwo
 \else \expandafter \@secondoftwo
 \fi
}%
\providecommand \natexlab [1]{#1}%
\providecommand \enquote  [1]{``#1''}%
\providecommand \bibnamefont  [1]{#1}%
\providecommand \bibfnamefont [1]{#1}%
\providecommand \citenamefont [1]{#1}%
\providecommand \href@noop [0]{\@secondoftwo}%
\providecommand \href [0]{\begingroup \@sanitize@url \@href}%
\providecommand \@href[1]{\@@startlink{#1}\@@href}%
\providecommand \@@href[1]{\endgroup#1\@@endlink}%
\providecommand \@sanitize@url [0]{\catcode `\\12\catcode `\$12\catcode
  `\&12\catcode `\#12\catcode `\^12\catcode `\_12\catcode `\%12\relax}%
\providecommand \@@startlink[1]{}%
\providecommand \@@endlink[0]{}%
\providecommand \url  [0]{\begingroup\@sanitize@url \@url }%
\providecommand \@url [1]{\endgroup\@href {#1}{\urlprefix }}%
\providecommand \urlprefix  [0]{URL }%
\providecommand \Eprint [0]{\href }%
\providecommand \doibase [0]{https://doi.org/}%
\providecommand \selectlanguage [0]{\@gobble}%
\providecommand \bibinfo  [0]{\@secondoftwo}%
\providecommand \bibfield  [0]{\@secondoftwo}%
\providecommand \translation [1]{[#1]}%
\providecommand \BibitemOpen [0]{}%
\providecommand \bibitemStop [0]{}%
\providecommand \bibitemNoStop [0]{.\EOS\space}%
\providecommand \EOS [0]{\spacefactor3000\relax}%
\providecommand \BibitemShut  [1]{\csname bibitem#1\endcsname}%
\let\auto@bib@innerbib\@empty
\bibitem [{\citenamefont {Ostrogradsky}(1850)}]{Ostrogradsky:1850fid}%
  \BibitemOpen
  \bibfield  {author} {\bibinfo {author} {\bibfnamefont {M.}~\bibnamefont
  {Ostrogradsky}},\ }\bibfield  {title} {\bibinfo {title} {{M\'emoires sur les
  \'equations diff\'erentielles, relatives au probl\`eme des
  isop\'erim\`etres}},\ }\href@noop {} {\bibfield  {journal} {\bibinfo
  {journal} {Mem. Acad. St. Petersbourg}\ }\textbf {\bibinfo {volume} {6}},\
  \bibinfo {pages} {385} (\bibinfo {year} {1850})}\BibitemShut {NoStop}%
\bibitem [{\citenamefont {Woodard}(2015)}]{Woodard:2015zca}%
  \BibitemOpen
  \bibfield  {author} {\bibinfo {author} {\bibfnamefont {R.~P.}\ \bibnamefont
  {Woodard}},\ }\bibfield  {title} {\bibinfo {title} {{Ostrogradsky's theorem
  on Hamiltonian instability}},\ }\href
  {https://doi.org/10.4249/scholarpedia.32243} {\bibfield  {journal} {\bibinfo
  {journal} {Scholarpedia}\ }\textbf {\bibinfo {volume} {10}},\ \bibinfo
  {pages} {32243} (\bibinfo {year} {2015})},\ \Eprint
  {https://arxiv.org/abs/1506.02210} {arXiv:1506.02210 [hep-th]} \BibitemShut
  {NoStop}%
\bibitem [{\citenamefont {Zumalac\'arregui}\ and\ \citenamefont
  {Garc\'\i{}a-Bellido}(2014)}]{Zumalacarregui:2013pma}%
  \BibitemOpen
  \bibfield  {author} {\bibinfo {author} {\bibfnamefont {M.}~\bibnamefont
  {Zumalac\'arregui}}\ and\ \bibinfo {author} {\bibfnamefont {J.}~\bibnamefont
  {Garc\'\i{}a-Bellido}},\ }\bibfield  {title} {\bibinfo {title} {{Transforming
  gravity: from derivative couplings to matter to second-order scalar-tensor
  theories beyond the Horndeski Lagrangian}},\ }\href
  {https://doi.org/10.1103/PhysRevD.89.064046} {\bibfield  {journal} {\bibinfo
  {journal} {Phys. Rev. D}\ }\textbf {\bibinfo {volume} {89}},\ \bibinfo
  {pages} {064046} (\bibinfo {year} {2014})},\ \Eprint
  {https://arxiv.org/abs/1308.4685} {arXiv:1308.4685 [gr-qc]} \BibitemShut
  {NoStop}%
\bibitem [{\citenamefont {Gleyzes}\ \emph
  {et~al.}(2015{\natexlab{a}})\citenamefont {Gleyzes}, \citenamefont
  {Langlois}, \citenamefont {Piazza},\ and\ \citenamefont
  {Vernizzi}}]{Gleyzes:2014dya}%
  \BibitemOpen
  \bibfield  {author} {\bibinfo {author} {\bibfnamefont {J.}~\bibnamefont
  {Gleyzes}}, \bibinfo {author} {\bibfnamefont {D.}~\bibnamefont {Langlois}},
  \bibinfo {author} {\bibfnamefont {F.}~\bibnamefont {Piazza}},\ and\ \bibinfo
  {author} {\bibfnamefont {F.}~\bibnamefont {Vernizzi}},\ }\bibfield  {title}
  {\bibinfo {title} {{Healthy theories beyond Horndeski}},\ }\href
  {https://doi.org/10.1103/PhysRevLett.114.211101} {\bibfield  {journal}
  {\bibinfo  {journal} {Phys. Rev. Lett.}\ }\textbf {\bibinfo {volume} {114}},\
  \bibinfo {pages} {211101} (\bibinfo {year} {2015}{\natexlab{a}})},\ \Eprint
  {https://arxiv.org/abs/1404.6495} {arXiv:1404.6495 [hep-th]} \BibitemShut
  {NoStop}%
\bibitem [{\citenamefont {Gleyzes}\ \emph
  {et~al.}(2015{\natexlab{b}})\citenamefont {Gleyzes}, \citenamefont
  {Langlois}, \citenamefont {Piazza},\ and\ \citenamefont
  {Vernizzi}}]{Gleyzes:2014qga}%
  \BibitemOpen
  \bibfield  {author} {\bibinfo {author} {\bibfnamefont {J.}~\bibnamefont
  {Gleyzes}}, \bibinfo {author} {\bibfnamefont {D.}~\bibnamefont {Langlois}},
  \bibinfo {author} {\bibfnamefont {F.}~\bibnamefont {Piazza}},\ and\ \bibinfo
  {author} {\bibfnamefont {F.}~\bibnamefont {Vernizzi}},\ }\bibfield  {title}
  {\bibinfo {title} {{Exploring gravitational theories beyond Horndeski}},\
  }\href {https://doi.org/10.1088/1475-7516/2015/02/018} {\bibfield  {journal}
  {\bibinfo  {journal} {JCAP}\ }\textbf {\bibinfo {volume} {02}},\ \bibinfo
  {pages} {018}},\ \Eprint {https://arxiv.org/abs/1408.1952} {arXiv:1408.1952
  [astro-ph.CO]} \BibitemShut {NoStop}%
\bibitem [{\citenamefont {Langlois}\ and\ \citenamefont
  {Noui}(2016)}]{Langlois:2015cwa}%
  \BibitemOpen
  \bibfield  {author} {\bibinfo {author} {\bibfnamefont {D.}~\bibnamefont
  {Langlois}}\ and\ \bibinfo {author} {\bibfnamefont {K.}~\bibnamefont
  {Noui}},\ }\bibfield  {title} {\bibinfo {title} {{Degenerate higher
  derivative theories beyond Horndeski: evading the Ostrogradski
  instability}},\ }\href {https://doi.org/10.1088/1475-7516/2016/02/034}
  {\bibfield  {journal} {\bibinfo  {journal} {JCAP}\ }\textbf {\bibinfo
  {volume} {02}},\ \bibinfo {pages} {034}},\ \Eprint
  {https://arxiv.org/abs/1510.06930} {arXiv:1510.06930 [gr-qc]} \BibitemShut
  {NoStop}%
\bibitem [{\citenamefont {Crisostomi}\ \emph {et~al.}(2016)\citenamefont
  {Crisostomi}, \citenamefont {Koyama},\ and\ \citenamefont
  {Tasinato}}]{Crisostomi:2016czh}%
  \BibitemOpen
  \bibfield  {author} {\bibinfo {author} {\bibfnamefont {M.}~\bibnamefont
  {Crisostomi}}, \bibinfo {author} {\bibfnamefont {K.}~\bibnamefont {Koyama}},\
  and\ \bibinfo {author} {\bibfnamefont {G.}~\bibnamefont {Tasinato}},\
  }\bibfield  {title} {\bibinfo {title} {{Extended Scalar-Tensor Theories of
  Gravity}},\ }\href {https://doi.org/10.1088/1475-7516/2016/04/044} {\bibfield
   {journal} {\bibinfo  {journal} {JCAP}\ }\textbf {\bibinfo {volume} {04}},\
  \bibinfo {pages} {044}},\ \Eprint {https://arxiv.org/abs/1602.03119}
  {arXiv:1602.03119 [hep-th]} \BibitemShut {NoStop}%
\bibitem [{\citenamefont {Ben~Achour}\ \emph
  {et~al.}(2016{\natexlab{a}})\citenamefont {Ben~Achour}, \citenamefont
  {Langlois},\ and\ \citenamefont {Noui}}]{BenAchour:2016cay}%
  \BibitemOpen
  \bibfield  {author} {\bibinfo {author} {\bibfnamefont {J.}~\bibnamefont
  {Ben~Achour}}, \bibinfo {author} {\bibfnamefont {D.}~\bibnamefont
  {Langlois}},\ and\ \bibinfo {author} {\bibfnamefont {K.}~\bibnamefont
  {Noui}},\ }\bibfield  {title} {\bibinfo {title} {{Degenerate higher order
  scalar-tensor theories beyond Horndeski and disformal transformations}},\
  }\href {https://doi.org/10.1103/PhysRevD.93.124005} {\bibfield  {journal}
  {\bibinfo  {journal} {Phys. Rev. D}\ }\textbf {\bibinfo {volume} {93}},\
  \bibinfo {pages} {124005} (\bibinfo {year} {2016}{\natexlab{a}})},\ \Eprint
  {https://arxiv.org/abs/1602.08398} {arXiv:1602.08398 [gr-qc]} \BibitemShut
  {NoStop}%
\bibitem [{\citenamefont {Ben~Achour}\ \emph
  {et~al.}(2016{\natexlab{b}})\citenamefont {Ben~Achour}, \citenamefont
  {Crisostomi}, \citenamefont {Koyama}, \citenamefont {Langlois}, \citenamefont
  {Noui},\ and\ \citenamefont {Tasinato}}]{BenAchour:2016fzp}%
  \BibitemOpen
  \bibfield  {author} {\bibinfo {author} {\bibfnamefont {J.}~\bibnamefont
  {Ben~Achour}}, \bibinfo {author} {\bibfnamefont {M.}~\bibnamefont
  {Crisostomi}}, \bibinfo {author} {\bibfnamefont {K.}~\bibnamefont {Koyama}},
  \bibinfo {author} {\bibfnamefont {D.}~\bibnamefont {Langlois}}, \bibinfo
  {author} {\bibfnamefont {K.}~\bibnamefont {Noui}},\ and\ \bibinfo {author}
  {\bibfnamefont {G.}~\bibnamefont {Tasinato}},\ }\bibfield  {title} {\bibinfo
  {title} {{Degenerate higher order scalar-tensor theories beyond Horndeski up
  to cubic order}},\ }\href {https://doi.org/10.1007/JHEP12(2016)100}
  {\bibfield  {journal} {\bibinfo  {journal} {JHEP}\ }\textbf {\bibinfo
  {volume} {12}},\ \bibinfo {pages} {100}},\ \Eprint
  {https://arxiv.org/abs/1608.08135} {arXiv:1608.08135 [hep-th]} \BibitemShut
  {NoStop}%
\bibitem [{\citenamefont {Horndeski}(1974)}]{Horndeski:1974wa}%
  \BibitemOpen
  \bibfield  {author} {\bibinfo {author} {\bibfnamefont {G.~W.}\ \bibnamefont
  {Horndeski}},\ }\bibfield  {title} {\bibinfo {title} {{Second-order
  scalar-tensor field equations in a four-dimensional space}},\ }\href
  {https://doi.org/10.1007/BF01807638} {\bibfield  {journal} {\bibinfo
  {journal} {Int. J. Theor. Phys.}\ }\textbf {\bibinfo {volume} {10}},\
  \bibinfo {pages} {363} (\bibinfo {year} {1974})}\BibitemShut {NoStop}%
\bibitem [{\citenamefont {Deffayet}\ \emph {et~al.}(2011)\citenamefont
  {Deffayet}, \citenamefont {Gao}, \citenamefont {Steer},\ and\ \citenamefont
  {Zahariade}}]{Deffayet:2011gz}%
  \BibitemOpen
  \bibfield  {author} {\bibinfo {author} {\bibfnamefont {C.}~\bibnamefont
  {Deffayet}}, \bibinfo {author} {\bibfnamefont {X.}~\bibnamefont {Gao}},
  \bibinfo {author} {\bibfnamefont {D.~A.}\ \bibnamefont {Steer}},\ and\
  \bibinfo {author} {\bibfnamefont {G.}~\bibnamefont {Zahariade}},\ }\bibfield
  {title} {\bibinfo {title} {{From k-essence to generalised Galileons}},\
  }\href {https://doi.org/10.1103/PhysRevD.84.064039} {\bibfield  {journal}
  {\bibinfo  {journal} {Phys. Rev. D}\ }\textbf {\bibinfo {volume} {84}},\
  \bibinfo {pages} {064039} (\bibinfo {year} {2011})},\ \Eprint
  {https://arxiv.org/abs/1103.3260} {arXiv:1103.3260 [hep-th]} \BibitemShut
  {NoStop}%
\bibitem [{\citenamefont {Kobayashi}\ \emph {et~al.}(2011)\citenamefont
  {Kobayashi}, \citenamefont {Yamaguchi},\ and\ \citenamefont
  {Yokoyama}}]{Kobayashi:2011nu}%
  \BibitemOpen
  \bibfield  {author} {\bibinfo {author} {\bibfnamefont {T.}~\bibnamefont
  {Kobayashi}}, \bibinfo {author} {\bibfnamefont {M.}~\bibnamefont
  {Yamaguchi}},\ and\ \bibinfo {author} {\bibfnamefont {J.}~\bibnamefont
  {Yokoyama}},\ }\bibfield  {title} {\bibinfo {title} {{Generalized
  G-inflation: Inflation with the most general second-order field equations}},\
  }\href {https://doi.org/10.1143/PTP.126.511} {\bibfield  {journal} {\bibinfo
  {journal} {Prog. Theor. Phys.}\ }\textbf {\bibinfo {volume} {126}},\ \bibinfo
  {pages} {511} (\bibinfo {year} {2011})},\ \Eprint
  {https://arxiv.org/abs/1105.5723} {arXiv:1105.5723 [hep-th]} \BibitemShut
  {NoStop}%
\bibitem [{\citenamefont {Langlois}(2019)}]{Langlois:2018dxi}%
  \BibitemOpen
  \bibfield  {author} {\bibinfo {author} {\bibfnamefont {D.}~\bibnamefont
  {Langlois}},\ }\bibfield  {title} {\bibinfo {title} {{Dark energy and
  modified gravity in degenerate higher-order scalar\textendash{}tensor (DHOST)
  theories: A review}},\ }\href {https://doi.org/10.1142/S0218271819420069}
  {\bibfield  {journal} {\bibinfo  {journal} {Int. J. Mod. Phys. D}\ }\textbf
  {\bibinfo {volume} {28}},\ \bibinfo {pages} {1942006} (\bibinfo {year}
  {2019})},\ \Eprint {https://arxiv.org/abs/1811.06271} {arXiv:1811.06271
  [gr-qc]} \BibitemShut {NoStop}%
\bibitem [{\citenamefont {Kobayashi}(2019)}]{Kobayashi:2019hrl}%
  \BibitemOpen
  \bibfield  {author} {\bibinfo {author} {\bibfnamefont {T.}~\bibnamefont
  {Kobayashi}},\ }\bibfield  {title} {\bibinfo {title} {{Horndeski theory and
  beyond: a review}},\ }\href {https://doi.org/10.1088/1361-6633/ab2429}
  {\bibfield  {journal} {\bibinfo  {journal} {Rept. Prog. Phys.}\ }\textbf
  {\bibinfo {volume} {82}},\ \bibinfo {pages} {086901} (\bibinfo {year}
  {2019})},\ \Eprint {https://arxiv.org/abs/1901.07183} {arXiv:1901.07183
  [gr-qc]} \BibitemShut {NoStop}%
\bibitem [{\citenamefont {Alinea}\ and\ \citenamefont
  {Kubota}(2021)}]{Alinea:2020laa}%
  \BibitemOpen
  \bibfield  {author} {\bibinfo {author} {\bibfnamefont {A.~L.}\ \bibnamefont
  {Alinea}}\ and\ \bibinfo {author} {\bibfnamefont {T.}~\bibnamefont
  {Kubota}},\ }\bibfield  {title} {\bibinfo {title} {{Transformation of
  primordial cosmological perturbations under the general extended disformal
  transformation}},\ }\href {https://doi.org/10.1142/S0218271821500577}
  {\bibfield  {journal} {\bibinfo  {journal} {Int. J. Mod. Phys. D}\ }\textbf
  {\bibinfo {volume} {30}},\ \bibinfo {pages} {2150057} (\bibinfo {year}
  {2021})},\ \Eprint {https://arxiv.org/abs/2005.12747} {arXiv:2005.12747
  [gr-qc]} \BibitemShut {NoStop}%
\bibitem [{\citenamefont {Minamitsuji}(2021)}]{Minamitsuji:2021dkf}%
  \BibitemOpen
  \bibfield  {author} {\bibinfo {author} {\bibfnamefont {M.}~\bibnamefont
  {Minamitsuji}},\ }\bibfield  {title} {\bibinfo {title} {{Generalized
  disformal invariance of cosmological perturbations with second-order field
  derivatives}},\ }\href {https://doi.org/10.1016/j.physletb.2021.136240}
  {\bibfield  {journal} {\bibinfo  {journal} {Phys. Lett. B}\ }\textbf
  {\bibinfo {volume} {816}},\ \bibinfo {pages} {136240} (\bibinfo {year}
  {2021})},\ \Eprint {https://arxiv.org/abs/2104.03662} {arXiv:2104.03662
  [gr-qc]} \BibitemShut {NoStop}%
\bibitem [{\citenamefont {Babichev}\ \emph {et~al.}(2022)\citenamefont
  {Babichev}, \citenamefont {Izumi}, \citenamefont {Tanahashi},\ and\
  \citenamefont {Yamaguchi}}]{Babichev:2021bim}%
  \BibitemOpen
  \bibfield  {author} {\bibinfo {author} {\bibfnamefont {E.}~\bibnamefont
  {Babichev}}, \bibinfo {author} {\bibfnamefont {K.}~\bibnamefont {Izumi}},
  \bibinfo {author} {\bibfnamefont {N.}~\bibnamefont {Tanahashi}},\ and\
  \bibinfo {author} {\bibfnamefont {M.}~\bibnamefont {Yamaguchi}},\ }\bibfield
  {title} {\bibinfo {title} {{Invertibility conditions for field
  transformations with derivatives: Toward extensions of disformal
  transformation with higher derivatives}},\ }\href
  {https://doi.org/10.1093/ptep/ptab151} {\bibfield  {journal} {\bibinfo
  {journal} {PTEP}\ }\textbf {\bibinfo {volume} {2022}},\ \bibinfo {pages}
  {013A01} (\bibinfo {year} {2022})},\ \Eprint
  {https://arxiv.org/abs/2109.00912} {arXiv:2109.00912 [hep-th]} \BibitemShut
  {NoStop}%
\bibitem [{\citenamefont {Babichev}\ \emph {et~al.}(2021)\citenamefont
  {Babichev}, \citenamefont {Izumi}, \citenamefont {Tanahashi},\ and\
  \citenamefont {Yamaguchi}}]{Babichev:2019twf}%
  \BibitemOpen
  \bibfield  {author} {\bibinfo {author} {\bibfnamefont {E.}~\bibnamefont
  {Babichev}}, \bibinfo {author} {\bibfnamefont {K.}~\bibnamefont {Izumi}},
  \bibinfo {author} {\bibfnamefont {N.}~\bibnamefont {Tanahashi}},\ and\
  \bibinfo {author} {\bibfnamefont {M.}~\bibnamefont {Yamaguchi}},\ }\bibfield
  {title} {\bibinfo {title} {{Invertible field transformations with
  derivatives: necessary and sufficient conditions}},\ }\href
  {https://doi.org/10.4310/ATMP.2021.v25.n2.a2} {\bibfield  {journal} {\bibinfo
   {journal} {Adv. Theor. Math. Phys.}\ }\textbf {\bibinfo {volume} {25}},\
  \bibinfo {pages} {309} (\bibinfo {year} {2021})},\ \Eprint
  {https://arxiv.org/abs/1907.12333} {arXiv:1907.12333 [hep-th]} \BibitemShut
  {NoStop}%
\bibitem [{\citenamefont {Takahashi}\ \emph {et~al.}(2022)\citenamefont
  {Takahashi}, \citenamefont {Motohashi},\ and\ \citenamefont
  {Minamitsuji}}]{Takahashi:2021ttd}%
  \BibitemOpen
  \bibfield  {author} {\bibinfo {author} {\bibfnamefont {K.}~\bibnamefont
  {Takahashi}}, \bibinfo {author} {\bibfnamefont {H.}~\bibnamefont
  {Motohashi}},\ and\ \bibinfo {author} {\bibfnamefont {M.}~\bibnamefont
  {Minamitsuji}},\ }\bibfield  {title} {\bibinfo {title} {{Invertible disformal
  transformations with higher derivatives}},\ }\href
  {https://doi.org/10.1103/PhysRevD.105.024015} {\bibfield  {journal} {\bibinfo
   {journal} {Phys. Rev. D}\ }\textbf {\bibinfo {volume} {105}},\ \bibinfo
  {pages} {024015} (\bibinfo {year} {2022})},\ \Eprint
  {https://arxiv.org/abs/2111.11634} {arXiv:2111.11634 [gr-qc]} \BibitemShut
  {NoStop}%
\bibitem [{\citenamefont {Motohashi}\ \emph
  {et~al.}(2018{\natexlab{a}})\citenamefont {Motohashi}, \citenamefont
  {Suyama},\ and\ \citenamefont {Yamaguchi}}]{Motohashi:2017eya}%
  \BibitemOpen
  \bibfield  {author} {\bibinfo {author} {\bibfnamefont {H.}~\bibnamefont
  {Motohashi}}, \bibinfo {author} {\bibfnamefont {T.}~\bibnamefont {Suyama}},\
  and\ \bibinfo {author} {\bibfnamefont {M.}~\bibnamefont {Yamaguchi}},\
  }\bibfield  {title} {\bibinfo {title} {{Ghost-free theory with third-order
  time derivatives}},\ }\href {https://doi.org/10.7566/JPSJ.87.063401}
  {\bibfield  {journal} {\bibinfo  {journal} {J. Phys. Soc. Jap.}\ }\textbf
  {\bibinfo {volume} {87}},\ \bibinfo {pages} {063401} (\bibinfo {year}
  {2018}{\natexlab{a}})},\ \Eprint {https://arxiv.org/abs/1711.08125}
  {arXiv:1711.08125 [hep-th]} \BibitemShut {NoStop}%
\bibitem [{\citenamefont {Motohashi}\ \emph
  {et~al.}(2018{\natexlab{b}})\citenamefont {Motohashi}, \citenamefont
  {Suyama},\ and\ \citenamefont {Yamaguchi}}]{Motohashi:2018pxg}%
  \BibitemOpen
  \bibfield  {author} {\bibinfo {author} {\bibfnamefont {H.}~\bibnamefont
  {Motohashi}}, \bibinfo {author} {\bibfnamefont {T.}~\bibnamefont {Suyama}},\
  and\ \bibinfo {author} {\bibfnamefont {M.}~\bibnamefont {Yamaguchi}},\
  }\bibfield  {title} {\bibinfo {title} {{Ghost-free theories with arbitrary
  higher-order time derivatives}},\ }\href
  {https://doi.org/10.1007/JHEP06(2018)133} {\bibfield  {journal} {\bibinfo
  {journal} {JHEP}\ }\textbf {\bibinfo {volume} {06}},\ \bibinfo {pages}
  {133}},\ \Eprint {https://arxiv.org/abs/1804.07990} {arXiv:1804.07990
  [hep-th]} \BibitemShut {NoStop}%
\bibitem [{\citenamefont {Bettoni}\ and\ \citenamefont
  {Liberati}(2013)}]{Bettoni:2013diz}%
  \BibitemOpen
  \bibfield  {author} {\bibinfo {author} {\bibfnamefont {D.}~\bibnamefont
  {Bettoni}}\ and\ \bibinfo {author} {\bibfnamefont {S.}~\bibnamefont
  {Liberati}},\ }\bibfield  {title} {\bibinfo {title} {{Disformal invariance of
  second order scalar-tensor theories: Framing the Horndeski action}},\ }\href
  {https://doi.org/10.1103/PhysRevD.88.084020} {\bibfield  {journal} {\bibinfo
  {journal} {Phys. Rev. D}\ }\textbf {\bibinfo {volume} {88}},\ \bibinfo
  {pages} {084020} (\bibinfo {year} {2013})},\ \Eprint
  {https://arxiv.org/abs/1306.6724} {arXiv:1306.6724 [gr-qc]} \BibitemShut
  {NoStop}%
\bibitem [{\citenamefont {Bekenstein}(1993)}]{Bekenstein:1992pj}%
  \BibitemOpen
  \bibfield  {author} {\bibinfo {author} {\bibfnamefont {J.~D.}\ \bibnamefont
  {Bekenstein}},\ }\bibfield  {title} {\bibinfo {title} {{The Relation between
  physical and gravitational geometry}},\ }\href
  {https://doi.org/10.1103/PhysRevD.48.3641} {\bibfield  {journal} {\bibinfo
  {journal} {Phys. Rev. D}\ }\textbf {\bibinfo {volume} {48}},\ \bibinfo
  {pages} {3641} (\bibinfo {year} {1993})},\ \Eprint
  {https://arxiv.org/abs/gr-qc/9211017} {arXiv:gr-qc/9211017} \BibitemShut
  {NoStop}%
\bibitem [{\citenamefont {Deruelle}\ and\ \citenamefont
  {Sasaki}(2011)}]{Deruelle:2010ht}%
  \BibitemOpen
  \bibfield  {author} {\bibinfo {author} {\bibfnamefont {N.}~\bibnamefont
  {Deruelle}}\ and\ \bibinfo {author} {\bibfnamefont {M.}~\bibnamefont
  {Sasaki}},\ }\bibfield  {title} {\bibinfo {title} {{Conformal equivalence in
  classical gravity: the example of 'Veiled' General Relativity}},\ }\href
  {https://doi.org/10.1007/978-3-642-19760-4_23} {\bibfield  {journal}
  {\bibinfo  {journal} {Springer Proc. Phys.}\ }\textbf {\bibinfo {volume}
  {137}},\ \bibinfo {pages} {247} (\bibinfo {year} {2011})},\ \Eprint
  {https://arxiv.org/abs/1007.3563} {arXiv:1007.3563 [gr-qc]} \BibitemShut
  {NoStop}%
\bibitem [{\citenamefont {Deruelle}\ and\ \citenamefont
  {Rua}(2014)}]{Deruelle:2014zza}%
  \BibitemOpen
  \bibfield  {author} {\bibinfo {author} {\bibfnamefont {N.}~\bibnamefont
  {Deruelle}}\ and\ \bibinfo {author} {\bibfnamefont {J.}~\bibnamefont {Rua}},\
  }\bibfield  {title} {\bibinfo {title} {{Disformal Transformations, Veiled
  General Relativity and Mimetic Gravity}},\ }\href
  {https://doi.org/10.1088/1475-7516/2014/09/002} {\bibfield  {journal}
  {\bibinfo  {journal} {JCAP}\ }\textbf {\bibinfo {volume} {09}},\ \bibinfo
  {pages} {002}},\ \Eprint {https://arxiv.org/abs/1407.0825} {arXiv:1407.0825
  [gr-qc]} \BibitemShut {NoStop}%
\bibitem [{\citenamefont {Arroja}\ \emph {et~al.}(2015)\citenamefont {Arroja},
  \citenamefont {Bartolo}, \citenamefont {Karmakar},\ and\ \citenamefont
  {Matarrese}}]{Arroja:2015wpa}%
  \BibitemOpen
  \bibfield  {author} {\bibinfo {author} {\bibfnamefont {F.}~\bibnamefont
  {Arroja}}, \bibinfo {author} {\bibfnamefont {N.}~\bibnamefont {Bartolo}},
  \bibinfo {author} {\bibfnamefont {P.}~\bibnamefont {Karmakar}},\ and\
  \bibinfo {author} {\bibfnamefont {S.}~\bibnamefont {Matarrese}},\ }\bibfield
  {title} {\bibinfo {title} {{The two faces of mimetic Horndeski gravity:
  disformal transformations and Lagrange multiplier}},\ }\href
  {https://doi.org/10.1088/1475-7516/2015/09/051} {\bibfield  {journal}
  {\bibinfo  {journal} {JCAP}\ }\textbf {\bibinfo {volume} {09}},\ \bibinfo
  {pages} {051}},\ \Eprint {https://arxiv.org/abs/1506.08575} {arXiv:1506.08575
  [gr-qc]} \BibitemShut {NoStop}%
\bibitem [{\citenamefont {Dom\`enech}\ \emph {et~al.}(2015)\citenamefont
  {Dom\`enech}, \citenamefont {Mukohyama}, \citenamefont {Namba}, \citenamefont
  {Naruko}, \citenamefont {Saitou},\ and\ \citenamefont
  {Watanabe}}]{Domenech:2015tca}%
  \BibitemOpen
  \bibfield  {author} {\bibinfo {author} {\bibfnamefont {G.}~\bibnamefont
  {Dom\`enech}}, \bibinfo {author} {\bibfnamefont {S.}~\bibnamefont
  {Mukohyama}}, \bibinfo {author} {\bibfnamefont {R.}~\bibnamefont {Namba}},
  \bibinfo {author} {\bibfnamefont {A.}~\bibnamefont {Naruko}}, \bibinfo
  {author} {\bibfnamefont {R.}~\bibnamefont {Saitou}},\ and\ \bibinfo {author}
  {\bibfnamefont {Y.}~\bibnamefont {Watanabe}},\ }\bibfield  {title} {\bibinfo
  {title} {{Derivative-dependent metric transformation and physical degrees of
  freedom}},\ }\href {https://doi.org/10.1103/PhysRevD.92.084027} {\bibfield
  {journal} {\bibinfo  {journal} {Phys. Rev. D}\ }\textbf {\bibinfo {volume}
  {92}},\ \bibinfo {pages} {084027} (\bibinfo {year} {2015})},\ \Eprint
  {https://arxiv.org/abs/1507.05390} {arXiv:1507.05390 [hep-th]} \BibitemShut
  {NoStop}%
\bibitem [{\citenamefont {Takahashi}\ \emph {et~al.}(2017)\citenamefont
  {Takahashi}, \citenamefont {Motohashi}, \citenamefont {Suyama},\ and\
  \citenamefont {Kobayashi}}]{Takahashi:2017zgr}%
  \BibitemOpen
  \bibfield  {author} {\bibinfo {author} {\bibfnamefont {K.}~\bibnamefont
  {Takahashi}}, \bibinfo {author} {\bibfnamefont {H.}~\bibnamefont
  {Motohashi}}, \bibinfo {author} {\bibfnamefont {T.}~\bibnamefont {Suyama}},\
  and\ \bibinfo {author} {\bibfnamefont {T.}~\bibnamefont {Kobayashi}},\
  }\bibfield  {title} {\bibinfo {title} {{General invertible transformation and
  physical degrees of freedom}},\ }\href
  {https://doi.org/10.1103/PhysRevD.95.084053} {\bibfield  {journal} {\bibinfo
  {journal} {Phys. Rev. D}\ }\textbf {\bibinfo {volume} {95}},\ \bibinfo
  {pages} {084053} (\bibinfo {year} {2017})},\ \Eprint
  {https://arxiv.org/abs/1702.01849} {arXiv:1702.01849 [gr-qc]} \BibitemShut
  {NoStop}%
\bibitem [{\citenamefont {Jirou\v{s}ek}\ \emph {et~al.}(2022)\citenamefont
  {Jirou\v{s}ek}, \citenamefont {Shimada}, \citenamefont {Vikman},\ and\
  \citenamefont {Yamaguchi}}]{Jirousek:2022jhh}%
  \BibitemOpen
  \bibfield  {author} {\bibinfo {author} {\bibfnamefont {P.}~\bibnamefont
  {Jirou\v{s}ek}}, \bibinfo {author} {\bibfnamefont {K.}~\bibnamefont
  {Shimada}}, \bibinfo {author} {\bibfnamefont {A.}~\bibnamefont {Vikman}},\
  and\ \bibinfo {author} {\bibfnamefont {M.}~\bibnamefont {Yamaguchi}},\
  }\bibfield  {title} {\bibinfo {title} {{New Dynamical Degrees of Freedom from
  Invertible Transformations}},\ }\href@noop {} {\  (\bibinfo {year} {2022})},\
  \Eprint {https://arxiv.org/abs/2208.05951} {arXiv:2208.05951 [gr-qc]}
  \BibitemShut {NoStop}%
\bibitem [{\citenamefont {Deffayet}\ and\ \citenamefont
  {Garcia-Saenz}(2020)}]{Deffayet:2020ypa}%
  \BibitemOpen
  \bibfield  {author} {\bibinfo {author} {\bibfnamefont {C.}~\bibnamefont
  {Deffayet}}\ and\ \bibinfo {author} {\bibfnamefont {S.}~\bibnamefont
  {Garcia-Saenz}},\ }\bibfield  {title} {\bibinfo {title} {{Degeneracy, matter
  coupling, and disformal transformations in scalar-tensor theories}},\ }\href
  {https://doi.org/10.1103/PhysRevD.102.064037} {\bibfield  {journal} {\bibinfo
   {journal} {Phys. Rev. D}\ }\textbf {\bibinfo {volume} {102}},\ \bibinfo
  {pages} {064037} (\bibinfo {year} {2020})},\ \Eprint
  {https://arxiv.org/abs/2004.11619} {arXiv:2004.11619 [hep-th]} \BibitemShut
  {NoStop}%
\bibitem [{\citenamefont {De~Felice}\ \emph {et~al.}(2018)\citenamefont
  {De~Felice}, \citenamefont {Langlois}, \citenamefont {Mukohyama},
  \citenamefont {Noui},\ and\ \citenamefont {Wang}}]{DeFelice:2018ewo}%
  \BibitemOpen
  \bibfield  {author} {\bibinfo {author} {\bibfnamefont {A.}~\bibnamefont
  {De~Felice}}, \bibinfo {author} {\bibfnamefont {D.}~\bibnamefont {Langlois}},
  \bibinfo {author} {\bibfnamefont {S.}~\bibnamefont {Mukohyama}}, \bibinfo
  {author} {\bibfnamefont {K.}~\bibnamefont {Noui}},\ and\ \bibinfo {author}
  {\bibfnamefont {A.}~\bibnamefont {Wang}},\ }\bibfield  {title} {\bibinfo
  {title} {{Generalized instantaneous modes in higher-order scalar-tensor
  theories}},\ }\href {https://doi.org/10.1103/PhysRevD.98.084024} {\bibfield
  {journal} {\bibinfo  {journal} {Phys. Rev. D}\ }\textbf {\bibinfo {volume}
  {98}},\ \bibinfo {pages} {084024} (\bibinfo {year} {2018})},\ \Eprint
  {https://arxiv.org/abs/1803.06241} {arXiv:1803.06241 [hep-th]} \BibitemShut
  {NoStop}%
\bibitem [{\citenamefont {Takahashi}\ \emph {et~al.}(2023)\citenamefont
  {Takahashi}, \citenamefont {Minamitsuji},\ and\ \citenamefont
  {Motohashi}}]{Takahashi:2022mew}%
  \BibitemOpen
  \bibfield  {author} {\bibinfo {author} {\bibfnamefont {K.}~\bibnamefont
  {Takahashi}}, \bibinfo {author} {\bibfnamefont {M.}~\bibnamefont
  {Minamitsuji}},\ and\ \bibinfo {author} {\bibfnamefont {H.}~\bibnamefont
  {Motohashi}},\ }\bibfield  {title} {\bibinfo {title} {{Generalized disformal
  Horndeski theories: cosmological perturbations and consistent matter
  coupling}},\ }\href {https://doi.org/10.1093/ptep/ptac161} {\bibfield
  {journal} {\bibinfo  {journal} {PTEP}\ }\textbf {\bibinfo {volume} {2023}},\
  \bibinfo {pages} {013E01} (\bibinfo {year} {2023})},\ \Eprint
  {https://arxiv.org/abs/2209.02176} {arXiv:2209.02176 [gr-qc]} \BibitemShut
  {NoStop}%
\bibitem [{\citenamefont {Motohashi}\ \emph {et~al.}(2016)\citenamefont
  {Motohashi}, \citenamefont {Noui}, \citenamefont {Suyama}, \citenamefont
  {Yamaguchi},\ and\ \citenamefont {Langlois}}]{Motohashi:2016ftl}%
  \BibitemOpen
  \bibfield  {author} {\bibinfo {author} {\bibfnamefont {H.}~\bibnamefont
  {Motohashi}}, \bibinfo {author} {\bibfnamefont {K.}~\bibnamefont {Noui}},
  \bibinfo {author} {\bibfnamefont {T.}~\bibnamefont {Suyama}}, \bibinfo
  {author} {\bibfnamefont {M.}~\bibnamefont {Yamaguchi}},\ and\ \bibinfo
  {author} {\bibfnamefont {D.}~\bibnamefont {Langlois}},\ }\bibfield  {title}
  {\bibinfo {title} {{Healthy degenerate theories with higher derivatives}},\
  }\href {https://doi.org/10.1088/1475-7516/2016/07/033} {\bibfield  {journal}
  {\bibinfo  {journal} {JCAP}\ }\textbf {\bibinfo {volume} {07}},\ \bibinfo
  {pages} {033}},\ \Eprint {https://arxiv.org/abs/1603.09355} {arXiv:1603.09355
  [hep-th]} \BibitemShut {NoStop}%
\bibitem [{\citenamefont {Klein}\ and\ \citenamefont
  {Roest}(2016)}]{Klein:2016aiq}%
  \BibitemOpen
  \bibfield  {author} {\bibinfo {author} {\bibfnamefont {R.}~\bibnamefont
  {Klein}}\ and\ \bibinfo {author} {\bibfnamefont {D.}~\bibnamefont {Roest}},\
  }\bibfield  {title} {\bibinfo {title} {{Exorcising the Ostrogradsky ghost in
  coupled systems}},\ }\href {https://doi.org/10.1007/JHEP07(2016)130}
  {\bibfield  {journal} {\bibinfo  {journal} {JHEP}\ }\textbf {\bibinfo
  {volume} {07}},\ \bibinfo {pages} {130}},\ \Eprint
  {https://arxiv.org/abs/1604.01719} {arXiv:1604.01719 [hep-th]} \BibitemShut
  {NoStop}%
\end{thebibliography}
